\documentclass[11pt]{article}

\usepackage[margin=1in]{geometry}
\usepackage{lmodern,amsmath,amssymb}
\usepackage{tikz}
\usetikzlibrary{positioning, arrows.meta, calc}
\usepackage{booktabs}
\usepackage{hyperref}
\usepackage{natbib}
\usepackage{graphicx}
\usepackage{fancyvrb}
\usepackage{xcolor}
\usepackage{arxiv}

\newcommand{\orcidicon}[1]{\href{https://orcid.org/#1}{\mbox{\begin{tikzpicture}[baseline=-0.2em]
\fill[color={rgb,255:red,166;green,206;blue,57}] (0,0) circle (0.55em);
\node[white,font=\bfseries\scriptsize\sffamily] at (0,0) {iD};
\end{tikzpicture}}}}

\newcommand{\proglang}[1]{\textsf{#1}}
\newcommand{\pkg}[1]{\textbf{#1}}
\let\code\texttt

\DefineVerbatimEnvironment{CodeInput}{Verbatim}{fontsize=\small,xleftmargin=1em}
\DefineVerbatimEnvironment{CodeOutput}{Verbatim}{fontsize=\small,xleftmargin=1em,frame=leftline,framerule=0.4pt,rulecolor=\color{gray}}

\title{bioLeak: Leakage-Aware Modeling and Diagnostics\\
for Machine Learning in \proglang{R}}
\author{Sel\c{c}uk Korkmaz\,\orcidicon{0000-0003-4632-6850}\\
  \small Department of Biostatistics and Medical Informatics,
  Faculty of Medicine\\
  \small Trakya University, 22030 Edirne, T\"urkiye\\
  \small \texttt{selcukorkmaz@gmail.com}
}
\date{}

\begin{document}
\maketitle

\begin{abstract}
Data leakage remains a recurrent source of optimistic bias in biomedical machine
learning studies. Standard row-wise cross-validation and globally estimated
preprocessing steps are often inappropriate for data with repeated
measurements, study-level heterogeneity, batch effects, or temporal
dependencies. This paper describes \pkg{bioLeak}, an \proglang{R} package for
constructing leakage-aware resampling workflows and for auditing fitted models
for common leakage mechanisms. The package provides leakage-aware split
construction, train-fold-only preprocessing, cross-validated model fitting,
nested hyperparameter tuning, post hoc leakage audits, and HTML reporting.
The implementation supports binary classification,
multiclass classification, regression, and survival analysis, with
task-specific metrics and S4 containers for splits, fits, audits, and
inflation summaries. The simulation artifacts show how apparent
performance changes under controlled leakage mechanisms, and the case study
illustrates how guarded and leaky pipelines can yield materially different
conclusions on multi-study transcriptomic data. The emphasis throughout is on
software design, reproducible workflows, and interpretation of diagnostic
output.
\end{abstract}


\bigskip

\section{Introduction}\label{sec:intro}

Cross-validated performance estimates are often interpreted as evidence that a predictive model will generalize to future biomedical data. That interpretation depends on assumptions that are frequently violated in practice: observations must be exchangeable with respect to the resampling design, preprocessing must be estimated without access to evaluation data, and the outcome must not be recoverable from proxy variables that would be unavailable at prediction time. Violations of these assumptions are commonly described as data leakage. Leakage is not a minor implementation detail. It affects whether estimated discrimination, calibration, and downstream biological interpretation are credible. This concern has been recognized for decades across statistical learning and biomedical prediction, from early discussions of selection bias in microarray studies \citep{ambroise2002} and broader treatments of leakage in data mining \citep{kaufman2012} to warnings about genomic prediction pitfalls \citep{simon2003, subramanian2013} and recent accounts linking leakage to wider reproducibility failures in machine-learning-based science \citep{kapoor2023}. Closely related evaluation bias also arises when model selection is not nested appropriately within resampling \citep{varma2006, cawley2010}.

The problem is especially acute in biomedical applications. A single study may include repeated measurements from the same participant, technical replicates, laboratory batches, study-site effects, or longitudinal samples collected over time. In such settings, ordinary row-wise cross-validation can place correlated samples in both training and test folds. Likewise, global imputation, normalization, filtering, or feature screening performed before resampling can transmit information from assessment data into the training pipeline. These choices can materially inflate apparent performance while leaving little trace in a conventional model summary, a practical concern in health-care machine learning where evaluation, implementation, and deployment constraints are already difficult to align \citep{beam2018, sendak2020}.

Recent methodological work has further clarified the scale and heterogeneity of the problem. \citet{kapoor2023} described leakage as one contributor to broader reproducibility failures in machine-learning-based science. In neuroimaging, \citet{rosenblatt2024} showed that repeated-subject leakage and feature-selection leakage can substantially inflate predictive performance, while \citet{vandemortel2025} reported that leakage-compatible practices were common enough to affect meta-analytic conclusions in psychiatric imaging. These studies suggest that leakage cannot be addressed by a single generic rule. It is heterogeneous, domain specific, and tightly tied to evaluation design.

Existing modeling frameworks provide much of the infrastructure needed for modern predictive analysis, but they do not by themselves define valid leakage boundaries. In \proglang{R} \citep{R}, \pkg{caret}, \pkg{tidymodels}, and \pkg{mlr3} support flexible resampling, preprocessing, and model-fitting workflows \citep{kuhn2008, kuhn2020, lang2019, bischl2024}. These frameworks can be used safely, but they rely on the analyst to encode resampling units, grouping variables, temporal structure, and fold-wise preprocessing correctly. In biomedical applications, where rows are often nested within subjects, batches, studies, or time windows, the central difficulty is not only fitting a model but specifying which observations may legitimately share information under the intended prediction setting.

\pkg{bioLeak} was developed to make those leakage boundaries explicit and operational within a single workflow. The package addresses two related tasks. The first is prevention: constructing split plans and preprocessing pipelines that reduce the risk of train--test contamination. The second is diagnosis: auditing fitted resampling results for patterns that are compatible with leakage or other structured sources of spurious signal and that may indicate inflated performance. Concretely, \pkg{bioLeak} provides leakage-aware split planning, overlap checking, guarded preprocessing, resampling-based model fitting, tuning where supported, and post hoc audit utilities within a unified \proglang{R} interface. The package is intended for practical use with tabular biomedical data and \code{SummarizedExperiment} objects, and it interoperates with widely used \proglang{R} infrastructure including \pkg{parsnip}, \pkg{recipes}, \pkg{workflows}, \pkg{yardstick}, and \pkg{rsample}.

The contribution of \pkg{bioLeak} is therefore not a new learner. Rather, it is a software framework for representing leakage-aware evaluation design as a first-class object, enforcing key train--test separation rules during preprocessing and resampling, and attaching audit procedures to fitted workflows so that suspicious performance inflation can be examined explicitly rather than inferred informally after the fact. This combination is particularly useful in biomedical settings, where dependency structure is often known in principle but inconsistently encoded in routine modeling practice.

The remainder of the manuscript is organized as follows. Section~\ref{sec:leakage} summarizes the leakage mechanisms that motivate the package. Section~\ref{sec:overview} describes the main software components. Section~\ref{sec:workflow} presents a reproducible workflow based on the package functions. Section~\ref{sec:dlsi} introduces the Delta Leakage Sensitivity Index ($\Delta_{\mathrm{LSI}}$) framework for quantifying performance inflation between leaky and guarded pipelines. Section~\ref{sec:simulation} reports simulation results, Section~\ref{sec:casestudy} presents a multi-study transcriptomic case study, and Section~\ref{sec:implementation} discusses the package implementation. Section~\ref{sec:discussion} addresses limitations and future directions.

\section{Data Leakage in Biomedical Machine Learning}\label{sec:leakage}
\subsection{Preprocessing leakage}

Preprocessing leakage occurs when quantities used to transform predictors are estimated using information from outside the training fold, rather than being learned strictly within each resampling split. Common examples include missing-value imputation, normalization, feature filtering, and feature selection applied before resampling or estimated on the full dataset. \citet{moscovich2022} showed that even unsupervised preprocessing can bias cross-validation when transformation parameters are estimated outside the resampling loop. This problem is common in biomedical pipelines, where normalization and predictor filtering are often treated as preliminary data-cleaning steps rather than as components of the modeling procedure. In genomics, the same issue is often described as feature-selection bias: selecting genes before resampling can produce substantially inflated performance estimates even when the downstream classifier itself is simple \citep{ambroise2002, simon2003, subramanian2013}.

\subsection{Dependent samples and subject overlap}

Many biomedical datasets contain repeated measures, multi-region samples from one patient, family-related observations, or longitudinal observations from the same individual. If row-wise resampling is applied, records from the same underlying unit can appear in both training and test folds. In that case, the model may partially recognize subject identity or other unit-specific structure rather than the biological signal of interest. This risk is especially high when outcomes are stable within individuals or when repeated measurements are closely spaced and therefore highly correlated. More generally, this problem belongs to the broader class of structured resampling failures discussed by \citet{roberts2017}, \citet{wenger2012}, and \citet{valavi2019}, in which naive row-wise fold assignment ignores dependencies induced by grouping, time, space, or hierarchy.

\subsection{Batch and study confounding}

High-throughput molecular assays frequently exhibit batch effects associated with laboratory, plate, scanner, platform, or study site. If outcome labels are unevenly distributed across these technical groups, a model can use batch as a surrogate for outcome. \citet{soneson2014} and \citet{hamdan2023} showed that such confounding, as well as attempts to remove it improperly, can bias cross-validation-based performance estimates. In multi-study settings, however, the intended validation target is critical: if one study is held out at a time for external validation, perfect fold--study association is expected and should not be interpreted as a software error. More generally, valid resampling must respect the dependence structure that is relevant to the intended deployment setting \citep{roberts2017}.

\subsection{Temporal look-ahead leakage}

When data have a temporal order, random splitting can allow the training set to use information that would not have been available at the time predictions are made for the test set. The most direct example is an explicit future variable, such as a subsequent laboratory measurement or a later event summary, included as a predictor. A subtler form arises when preprocessing or feature construction uses information estimated from future observations. Proper time-aware resampling therefore requires chronological ordering and, in some settings, additional separation between training and evaluation periods to prevent temporal overlap. More generally, evaluation is meaningful only when the data split matches the prediction task that the model is intended to support \citep{stone1974, arlot2010, hastie2009}.

\subsection{Target leakage and duplicate samples}

Target leakage occurs when a predictor carries direct information about the outcome definition. In electronic health records, this may take the form of a billing code, procedure, or medication that is downstream of the diagnosis. In molecular or public expression datasets, it may arise through label-derived features, transformed outcome summaries, or variables created using information unavailable at prediction time. Duplicate and near-duplicate samples are related but distinct: they do not need to encode the outcome explicitly, yet a nearly identical training sample can make a test sample artificially easy to predict. Because these problems can persist even when resampling appears well designed, they are natural targets for post hoc diagnostics. Near-duplicate detection inherits standard design choices from record linkage and similarity search---feature representation, matching rules, and similarity thresholds---rather than classical hypothesis testing \citep{winkler1999, christen2012, leskovec2014}.

\section{Overview of the \pkg{bioLeak} Package}\label{sec:overview}

The package is organized around four design goals: leakage-aware modeling,
reproducible workflows, diagnostic auditing, and compatibility with common \proglang{R}
modeling tools. The goal is to evaluate whether reported performance may be inflated by leakage or other structured sources of spurious signal.

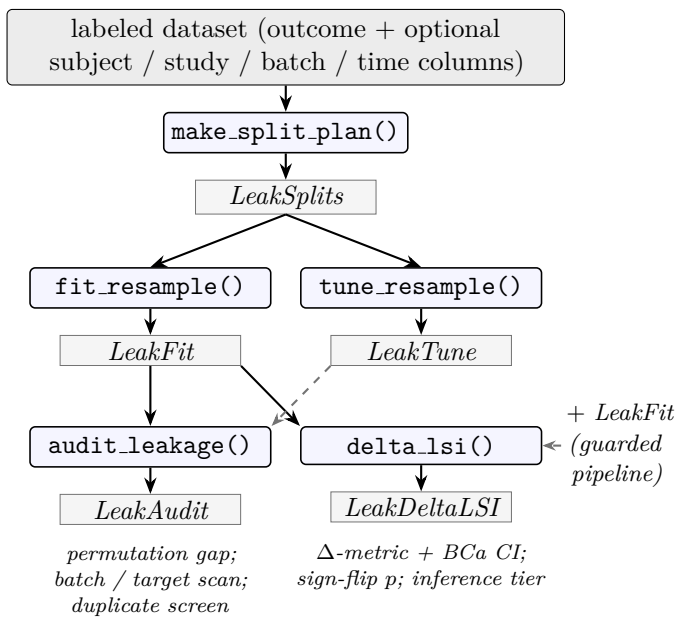
\begin{figure}[tbp]
\centering
\begin{tikzpicture}[
  node distance=0.35cm and 0.7cm,
  input/.style={draw, rectangle, rounded corners=2pt, align=center,
                fill=gray!15, minimum height=0.5cm, font=\small,
                inner sep=3pt, text width=7.2cm},
  fn/.style={draw, rectangle, rounded corners=2pt, align=center,
             fill=blue!4, minimum height=0.5cm, font=\ttfamily\small,
             inner sep=3pt, minimum width=3.2cm, line width=0.6pt},
  obj/.style={draw, rectangle, align=center,
              fill=gray!8, minimum height=0.4cm, font=\itshape\small,
              inner sep=2pt, minimum width=2.4cm, line width=0.3pt,
              draw=black!55},
  annot/.style={font=\footnotesize\itshape, align=left, text width=5cm},
  outannot/.style={font=\scriptsize\itshape, align=center, text width=3.4cm},
  sidein/.style={font=\footnotesize\itshape, align=left, text width=2.4cm},
  arr/.style={-{Stealth[length=2.2mm]}, thick},
  dasharr/.style={-{Stealth[length=1.8mm]}, thick, dashed, draw=black!55}
]

%% Main spine
\node[input] (data) {labeled dataset (outcome + optional\\ subject / study / batch / time columns)};
\node[fn, below=of data] (split) {make\_split\_plan()};
\node[obj, below=of split] (splits) {LeakSplits};

%% Parallel fitting paths: fit_resample (left) and tune_resample (right)
\node[fn, below=0.7cm of splits, xshift=-1.8cm] (fit) {fit\_resample()};
\node[fn, below=0.7cm of splits, xshift=1.8cm] (tune) {tune\_resample()};
\node[obj, below=of fit] (lfit) {LeakFit};
\node[obj, below=of tune] (ltune) {LeakTune};

%% Audit and delta_lsi below fitting layer
\node[fn, below=0.8cm of lfit] (audit) {audit\_leakage()};
\node[fn, below=0.8cm of ltune] (dlsi) {delta\_lsi()};

%% Terminal objects
\node[obj, below=of audit] (laudit) {LeakAudit};
\node[obj, below=of dlsi] (ldlsi) {LeakDeltaLSI};

%% Output annotations under terminal objects
\node[outannot, below=0.15cm of laudit]
  {permutation gap; batch / target scan; duplicate screen};
\node[outannot, below=0.15cm of ldlsi]
  {$\Delta$-metric + BCa CI; sign-flip $p$; inference tier};

%% Second-LeakFit side input feeding delta_lsi
\node[sidein, right=0.25cm of dlsi] (sideinput)
  {$+$ LeakFit \\ (guarded pipeline)};
\draw[dasharr] (sideinput.west) -- (dlsi.east);

%% Arrows
\draw[arr] (data) -- (split);
\draw[arr] (split) -- (splits);
\draw[arr] (splits.south) -- (fit.north);
\draw[arr] (splits.south) -- (tune.north);
\draw[arr] (fit) -- (lfit);
\draw[arr] (tune) -- (ltune);
%% LeakFit feeds both audit and delta_lsi; LeakTune feeds only audit
\draw[arr] (lfit) -- (audit);
\draw[arr] (lfit.south east) -- (dlsi.north west);
\draw[dasharr] (ltune.south west) -- (audit.north east);
\draw[arr] (audit) -- (laudit);
\draw[arr] (dlsi) -- (ldlsi);

\end{tikzpicture}
\caption{\pkg{bioLeak} leakage-aware modeling workflow. Rounded boxes
(\texttt{monospace}) are function calls; italic labels are the objects
they return. Solid arrows show the primary data flow; dashed arrows
indicate secondary inputs. \code{audit\_leakage()} accepts both
\code{LeakFit} and \code{LeakTune} objects; \code{delta\_lsi()}
compares two \code{LeakFit} objects (one from each pipeline).}
\label{fig:workflow}
\end{figure}

Figure~\ref{fig:workflow} summarizes the main package workflow from a labeled
dataset to fitted resampling results, audit summaries, and inflation
quantification.

Each stage produces an S4 object: \code{LeakSplits}, \code{LeakFit},
\code{LeakAudit}, and \code{LeakDeltaLSI}. These objects store the intermediate state needed for most downstream analysis and reporting steps.

The main workflow begins with \code{make\_split\_plan()}, which constructs an
S4 \code{LeakSplits} object. The split plan records the resampling mode,
fold-level indices or compact fold assignments, and metadata such as the
grouping variables, number of folds, number of repeats, and a reproducibility
hash. The split modes most relevant to the present manuscript are
\code{subject\_grouped}, \code{batch\_blocked},
\code{study\_loocv}, and \code{time\_series}. These modes address distinct
dependence structures rather than acting as interchangeable cross-validation
variants.

Model fitting is handled by \code{fit\_resample()}, which applies guarded
preprocessing independently within each fold, trains one or more learners, and
returns a \code{LeakFit} object. The fit object stores per-fold metrics,
aggregated summaries, audit metadata, predictions, preprocessing state,
learners, and information required for later refit-based audits. The function
accepts data frames, matrices, and \code{SummarizedExperiment} objects.
Learners may be specified through \pkg{parsnip}, \pkg{workflows}, legacy
character interfaces, or custom fit/predict pairs.

\code{tune\_resample()} extends this design to nested resampling. Outer folds
are defined by a \code{LeakSplits} object or an \pkg{rsample} structure
that already contains inner folds. Tuning results are returned as a
\code{LeakTune} object containing inner tuning outputs, outer-fold metrics,
selected hyperparameters, optional threshold selections for binary outcomes, and
an optional refitted final workflow. Survival tasks are currently excluded from
\code{tune\_resample()}.

Auditing is provided by \code{audit\_leakage()}, which returns a
\code{LeakAudit} object. The audit combines a permutation-gap analysis, fold
association tests for batch or study metadata, target-leakage scans on a
reference feature matrix, duplicate detection, and a compact mechanism-level
summary stored in the object metadata. \code{audit\_report()} renders these
results to a self-contained HTML report based on a bundled \proglang{R} Markdown template.
For direct inflation comparisons between a leaky and a guarded pipeline, the
package also provides \code{delta\_lsi()}, which returns a
\code{LeakDeltaLSI} object summarizing repeat-level performance differences,
robust point estimates, confidence intervals, and sign-flip randomization
tests.

Table~\ref{tab:components} lists the primary user-facing functions and their
roles in the workflow.

\begin{table}[tbp]
\centering
\caption{Primary functions in \pkg{bioLeak}.}
\label{tab:components}
\begin{tabular}{|p{0.24\linewidth}|p{0.23\linewidth}|p{0.41\linewidth}|}
\hline
\textbf{Stage} & \textbf{Function} & \textbf{Main output or role} \\
\hline
Split construction & \code{make\_split\_plan()} & Leakage-aware split plans for subject, batch, study, or time-ordered resampling. \\
\hline
Cross-validated fitting & \code{fit\_resample()} & Fold predictions, fold metrics, aggregated performance summaries, preprocessing state, and refit metadata. \\
\hline
Nested tuning & \code{tune\_resample()} & Nested cross-validation for hyperparameter tuning with leakage-aware outer and inner folds. \\
\hline
Leakage auditing & \code{audit\_leakage()} & Permutation-gap summaries, batch or study association tests, target scans, duplicate detection, and mechanism summaries. \\
\hline
Audit reporting & \code{audit\_report()} & Self-contained HTML summaries suitable for documentation and sharing. \\
\hline
$\Delta_{\mathrm{LSI}}$ comparison & \code{delta\_lsi()} & Repeat-level comparison of leaky and guarded pipelines, including robust inflation estimates and randomization-based inference. \\
\hline
Interoperability & \code{guard\_to\_recipe()}, \code{as\_rsample()} & Bridges to \pkg{recipes} and \pkg{rsample} workflows. \\
\hline
\end{tabular}
\end{table}

\section{Leakage-Aware Modeling Workflow}\label{sec:workflow}

This section illustrates a typical workflow using a small synthetic binary classification example. The same workflow structure can be adapted to multiclass, regression, and survival tasks, although the available metrics and some audit components differ across settings. The underlying evaluation logic follows standard resampling principles, but adapts them to structured biomedical data in which the relevant unit of independence may be subject, batch, study, or time rather than the individual row \citep{stone1974, kohavi1995, arlot2010}.

\subsection{Constructing a split plan}

The first step is to define the dependence structure that the resampling design should respect. In the example below, repeated measurements are nested within subject, so a subject-grouped split plan is used. The code mirrors the current package interface and can be adapted to \code{batch\_blocked}, \code{study\_loocv}, or \code{time\_series} settings by changing the resampling mode and the corresponding grouping or time-index columns.

\begin{CodeInput}
library(bioLeak)
library(parsnip)

## Simulate 40 subjects with 3 repeated measurements each
set.seed(1)
n_subject <- 40
rep_per_subject <- 3
n <- n_subject * rep_per_subject

subject <- rep(seq_len(n_subject), each = rep_per_subject)
batch <- rep(seq_len(6), length.out = n)
x1 <- rnorm(n)
x2 <- rnorm(n)
x3 <- rnorm(n)
subject_signal <- rnorm(n_subject)
outcome <- factor(
  ifelse(runif(n) < plogis(subject_signal[subject]), "case", "control"),
  levels = c("control", "case")
)

df <- data.frame(subject, batch, outcome, x1, x2, x3)

## Build a subject-grouped 5-fold split plan
splits <- make_split_plan(
  df,
  outcome = "outcome",
  mode = "subject_grouped",
  group = "subject",
  v = 5,
  stratify = TRUE,
  seed = 1
)

splits
\end{CodeInput}
\begin{CodeOutput}
LeakSplits object (mode = subject_grouped, v = 5, repeats = 1)
Outcome: outcome | Stratified: TRUE | Nested: FALSE
------------------------------------------------------
  repeat fold n_train n_test
       1    1      96     24
       1    2      96     24
       1    3      96     24
       1    4      96     24
       1    5      96     24
------------------------------------------------------
Total folds: 5 | Hash: a1b2c3d4e5f6
\end{CodeOutput}

\noindent
The split plan is created successfully and yields five subject-grouped folds of equal size. Because all repeated measurements from a given subject are assigned to the same fold, the resampling design prevents subject overlap between training and test sets. In this synthetic example, the object serves as the common evaluation structure for both the guarded and intentionally leaky fits considered below.

For large datasets, \code{compact = TRUE} stores fold assignments as a per-row fold vector rather than retaining explicit index lists for each fold, reducing memory use by roughly $v$-fold (e.g., from five copies of row indices to one integer vector).

\subsection{Defining guarded preprocessing}

\code{fit\_resample()} accepts a native guarded preprocessing specification
or a \pkg{recipes} recipe object. Under the native interface, each
preprocessing step is estimated on the training fold only and the resulting
parameters are then applied to the corresponding test fold. Supported
operations include imputation, normalization, filtering, and feature
selection. In the example below, median imputation and z-score normalization
are performed separately within each training fold.

\begin{CodeInput}
spec <- logistic_reg(mode = "classification") |>
  set_engine("glm")

fit_guarded <- fit_resample(
  df,
  outcome = "outcome",
  splits = splits,
  learner = spec,
  metrics = "auc",
  preprocess = list(
    impute = list(method = "median"),
    normalize = list(method = "zscore")
  ),
  refit = FALSE,
  seed = 1
)

summary(fit_guarded)
\end{CodeInput}
\begin{CodeOutput}
===========================
 bioLeak Model Fit Summary
===========================

Task: binomial
Outcome: outcome
Positive class: case
Learners: logistic_reg/glm
Total folds: 5
Fold status: 5 success, 0 skipped, 0 failed
Refit performed: No
Hash: 7a3f2e1b4c9d

Cross-validated metrics (mean +/- SD):
           learner auc_mean auc_sd auc_ci_lo auc_ci_hi
1 logistic_reg/glm    0.542  0.102      0.34      0.74
\end{CodeOutput}

\noindent
The guarded fit completes successfully across all five folds and yields near-chance discrimination (mean AUC 0.54), which is expected in this synthetic example because the predictors are mostly noise and no label-derived feature has been introduced. This fitted object therefore serves as the guarded baseline for the leaky comparator examined in the following subsection.

For survival analysis, the outcome is specified using time and event columns, and concordance-based evaluation is supported. For regression, common metrics include \code{rmse}, \code{mae}, and $R^2$. Survival support in the current version covers \code{fit\_resample()} and \code{delta\_lsi()} with concordance metrics; however, nested tuning, target scanning, and permutation-refit audits are not yet available for survival tasks. The worked example, simulation study, and case study in this paper all use binary classification; multiclass, regression, and survival workflows are not demonstrated here.

\subsection{Training leaky and guarded comparators}

The package is most informative when a guarded analysis is compared with a plausible naive workflow or with a deliberately constructed leakage scenario. Such comparisons are useful for sensitivity analysis and provide the basis for the $\Delta_{\mathrm{LSI}}$ procedure described in Section~\ref{sec:dlsi}. In the example below, a label-derived subject-level feature is added to create an intentionally leaky comparator while keeping the same grouped split structure. The resulting fitted object is used in the following subsection to quantify performance inflation relative to the guarded analysis.

\begin{CodeInput}
df_leaky <- within(df, {
  leak_subject <- ave(as.numeric(outcome == "case"), subject, FUN = mean)
})

fit_leaky <- fit_resample(
  df_leaky,
  outcome = "outcome",
  splits = splits,
  learner = spec,
  metrics = "auc",
  preprocess = list(
    impute = list(method = "none"),
    normalize = list(method = "none")
  ),
  refit = FALSE,
  seed = 1
)

summary(fit_leaky)
\end{CodeInput}
\begin{CodeOutput}
===========================
 bioLeak Model Fit Summary
===========================

Task: binomial
Outcome: outcome
Positive class: case
Learners: logistic_reg/glm
Total folds: 5
Fold status: 5 success, 0 skipped, 0 failed
Refit performed: No
Hash: h9bcac8d4

Cross-validated metrics (mean +/- SD):
           learner auc_mean auc_sd auc_ci_lo auc_ci_hi
1 logistic_reg/glm     0.92  0.054      0.82      1.00

Audit overview:
 fold n_train n_test          learner features_final
    1      96     24 logistic_reg/glm              5
    2      96     24 logistic_reg/glm              5
    3      96     24 logistic_reg/glm              5
    4      96     24 logistic_reg/glm              5
    5      96     24 logistic_reg/glm              5
\end{CodeOutput}

\noindent
The intentionally leaky comparator yields a markedly higher cross-validated AUC (0.92) than the guarded fit shown earlier (0.54). Because the grouped split structure is unchanged, this inflation is attributable to the label-derived subject-level feature rather than to a change in resampling design. The example therefore illustrates how leakage can inflate apparent predictive performance even when fold construction itself is appropriate.

\subsection{Auditing a fitted workflow}

After cross-validated predictions have been generated, \code{audit\_leakage()} can be used to summarize multiple forms of audit evidence from the fitted workflow and, optionally, from an aligned reference matrix \code{X\_ref}. The reference matrix enables target-leakage scans and near-duplicate detection. In grouped designs, the same interface can also be used to assess whether performance remains above chance under label permutation and whether fold assignments are associated with batch or study structure.

The call below requests $B = 50$ label permutations and sets \code{perm\_stratify = TRUE} so that the grouped split structure is respected when labels are shuffled. \code{X\_ref} receives only the three numeric predictor columns (excluding \code{subject}, \code{batch}, and \code{outcome}) to enable the target-leakage and duplicate scans; \code{batch\_cols} names the column tested for batch--outcome association; and \code{return\_perm = TRUE} retains the full permutation distribution in the returned object for further inspection.

\begin{CodeInput}
audit_guarded <- audit_leakage(
  fit_guarded,
  metric = "auc",
  B = 50,
  perm_stratify = TRUE,
  X_ref = df[, c("x1", "x2", "x3")],
  batch_cols = "batch",
  return_perm = TRUE
)

summary(audit_guarded)
\end{CodeInput}
\begin{CodeOutput}
==============================
 bioLeak Leakage Audit Summary
==============================

Task: binomial | Outcome: outcome | Splitting mode: subject_grouped
Hash: 7a3f2e1b4c9d | Folds: 5 | Repeats: 1

Label-Permutation Association Test:
  Method: fixed predictions
  Observed metric: 0.5417
  Permuted mean +/- SD: 0.5012 +/- 0.0389
  Gap: 0.0405 (larger gap = stronger non-random signal)

Batch / Study Association:
  batch (repeat 1): Chi^2 = 4.123 (df = 4), p = 0.3897

Target Leakage Scan:
  Features checked: 3 | Flagged (score >= 0.90): 0

Multivariate Target Scan: not available.

Near-Duplicate Samples:
  No near-duplicates detected.

Mechanism Risk Assessment:
  mechanism_class      flagged evidence
  subject_overlap      FALSE   OK: subject-grouped splits
  batch_confounded     FALSE   OK: batch Chi^2 not significant
  preprocessing_leak   FALSE   OK: guarded preprocessing
  target_leakage       FALSE   OK: no features flagged

Interpretation:
  Modest non-random signal. No leakage indicators flagged.
\end{CodeOutput}

\noindent
In this example, the audit detects a near-chance signal in the guarded fit (observed AUC $0.54$, permutation gap $0.04$ above the shuffled null) and finds no evidence of leakage: the batch--outcome $\chi^2$ test is non-significant ($p = 0.39$), no predictor exceeds the target-leakage threshold, and no near-duplicate samples are detected. The \emph{Multivariate Target Scan} is skipped because the synthetic example has too few predictors to build a meaningful principal-component model; it is enabled by default on richer feature matrices. The final \emph{Mechanism Risk Assessment} block consolidates the individual diagnostics into a per-mechanism verdict: each row corresponds to a leakage mechanism class (subject overlap, batch confounding, preprocessing leakage, target leakage) and is flagged only when at least one associated diagnostic crosses its decision threshold. None of the four classes are flagged, which is consistent with the clean guarded design.

The resulting \code{LeakAudit} object stores permutation summaries, feature-level target-association results, duplicate-detection output, and the audit settings used to generate them. The permutation module is diagnostic rather than dispositive: a positive gap suggests predictive signal beyond chance, but the remaining audit components are needed to judge whether that signal may be inflated by leakage. The audit object can additionally be rendered as a standalone HTML report via \code{audit\_report()}, which summarizes all components above for documentation or sharing.

\paragraph{Practical guidance on permutation calibration.}
By default, \code{fit\_resample()} stores the original data and learner
specification inside the \code{LeakFit} object (\code{store\_refit\_data
= TRUE}). When \code{audit\_leakage()} is subsequently called with the
default \code{perm\_refit = "auto"}, it detects this stored specification
and activates refit-based permutations, in which the model is retrained
under each label permutation. This yields a well-calibrated null
distribution with type~I error at the nominal level. The fixed-prediction
alternative (\code{perm\_refit = FALSE}), which the simulations in
Section~\ref{sec:simulation} use for computational reasons, is faster
but slightly liberal (empirical type~I error of $6$--$10.5\%$ depending
on $(n, p)$). Users who set \code{perm\_refit = FALSE} or
\code{store\_refit\_data = FALSE} should be aware of this trade-off;
the package emits an informational message when auto mode falls back to
fixed predictions.

\subsection{Putting the workflow together}

The previous subsections present each stage separately. In practice, the full
workflow is usually executed as a single script. The following code block
reuses the simulated data frame \code{df} from Section~\ref{sec:workflow} and
chains every stage into one end-to-end sequence.

\begin{CodeInput}
## Assumes df from Section 4.1 is in scope
dat <- df

## 1. Subject-grouped 5-fold split plan
plan <- make_split_plan(
  dat,
  outcome = "outcome",
  mode = "subject_grouped",
  group = "subject",
  v = 5,
  stratify = TRUE,
  seed = 1
)

## 2. Learner specification
spec <- logistic_reg(mode = "classification") |>
  set_engine("glm")

## 3. Guarded fit
fit_guarded <- fit_resample(
  dat,
  outcome = "outcome",
  splits = plan,
  learner = spec,
  metrics = "auc",
  refit = FALSE,
  seed = 1
)

## 4. Leaky comparator (label-derived subject-level feature)
dat_leaky <- within(dat, {
  leak_subject <- ave(as.numeric(outcome == "case"), subject, FUN = mean)
})

fit_leaky <- fit_resample(
  dat_leaky,
  outcome = "outcome",
  splits = plan,
  learner = spec,
  metrics = "auc",
  refit = FALSE,
  seed = 1
)

## 5. Audit the guarded fit
audit <- audit_leakage(
  fit_guarded,
  metric = "auc",
  B = 50,
  perm_stratify = TRUE,
  batch_cols = "batch",
  X_ref = dat[, c("x1", "x2", "x3")]
)

## 6. Quantify performance inflation (leaky vs guarded)
delta <- delta_lsi(
  fit_leaky = fit_leaky,
  fit_guarded = fit_guarded,
  metric = "auc",
  seed = 1
)

## 7. Render HTML audit report
report_file <- audit_report(
  audit,
  output_file = "bioLeak_audit_report.html",
  output_dir = tempdir(),
  quiet = TRUE
)
\end{CodeInput}

\noindent
The split plan defines the dependence structure, \code{fit\_resample()} trains the guarded and intentionally leaky pipelines under that design, \code{audit} stores diagnostic evidence for the guarded fit, \code{delta} summarizes performance inflation between the two analyses, and \code{audit\_report()} produces a shareable HTML summary.

\subsection{Leakage-safe hyperparameter tuning}

\code{tune\_resample()} performs nested cross-validation so that hyperparameter selection remains inside the outer resampling loop. The example below uses a tunable \pkg{glmnet} specification together with a nested split plan. This design avoids tuning on the same assessment data used for outer-loop evaluation.

The split plan is built with \code{nested = TRUE}, which attaches an inner resampling loop inside each of the three outer folds for hyperparameter search. In the \code{tune\_resample()} call, \code{grid = 5} requests five candidate penalty values explored by the inner loop, \code{selection = "one\_std\_err"} applies the one-standard-error rule to favor more parsimonious models when inner-fold performance differences are small, and \code{refit = TRUE} fits a final model on the full dataset using hyperparameters aggregated across all outer folds (median for numeric parameters, majority vote for categorical) rather than a single fold's optimum, so that the returned object carries a deployable model without introducing nested-CV selection bias.

\begin{CodeInput}
library(recipes)
library(tune)

## Nested subject-grouped split plan (inner loop for tuning,
## outer loop for evaluation)
nested_splits <- make_split_plan(
  df,
  outcome = "outcome",
  mode = "subject_grouped",
  group = "subject",
  v = 3,
  nested = TRUE,
  seed = 14
)

## Preprocessing recipe (applied inside each training fold)
rec <- recipe(outcome ~ x1 + x2 + x3, data = df) |>
  step_impute_median(all_numeric_predictors()) |>
  step_normalize(all_numeric_predictors())

## Tunable lasso specification
spec_tune <- logistic_reg(
  penalty = tune(),
  mixture = 1,
  mode = "classification"
) |>
  set_engine("glmnet")

## Nested tuning: penalty is selected on inner folds,
## evaluated on outer folds
tuned <- tune_resample(
  df,
  outcome = "outcome",
  splits = nested_splits,
  learner = spec_tune,
  preprocess = rec,
  grid = 5,
  metrics = c("auc", "accuracy"),
  selection = "one_std_err",
  refit = TRUE,
  seed = 14
)

summary(tuned)
\end{CodeInput}
\begin{CodeOutput}
============================
 bioLeak Tuning Summary
============================

Task: binomial
Outcome: outcome
Positive class: case
Tuning Grid: 5
Selection Rule: one_std_err (Metric: roc_auc)
Fold status: 3 success, 0 skipped, 0 failed
Refit performed: Yes
Outer Folds: 3 successful / 3 total

Outer Loop Metrics (mean +/- SD):
              learner accuracy_mean accuracy_sd roc_auc_mean roc_auc_sd
1 logistic_reg/glmnet         0.424       0.024        0.454      0.079
  accuracy_ci_lo accuracy_ci_hi roc_auc_ci_lo roc_auc_ci_hi
1          0.364          0.485         0.257         0.651

Best Parameters (First 5 Folds):
 penalty fold
   0.928    1
   0.838    2
   0.003    3
\end{CodeOutput}

\noindent
The resulting object stores outer-fold performance summaries together with the tuning results needed to identify the selected hyperparameter values under nested resampling. As expected on the synthetic noise predictors, the outer-fold \code{roc\_auc} is near chance (mean $0.45$, CI $0.26$--$0.65$), consistent with the guarded fit shown earlier (AUC $0.54$). The selected penalty also fluctuates across folds ($0.93$, $0.84$, $0.003$), which is typical when no real signal is present and illustrates why the nested design is needed: tuning on the same folds used for evaluation would lock in these fold-specific optima and report optimistic performance.

When a \code{recipe} or \code{workflow} is supplied, \pkg{bioLeak} applies guard checks to reduce the risk of preprocessing steps being estimated outside the intended fold structure. The package also provides \code{guard\_to\_recipe()}, which translates a guarded preprocessing specification into an approximate \pkg{recipes} pipeline so that users who prefer the \pkg{tidymodels} ecosystem can apply the same transformations; the function maps imputation, normalization, and filtering steps to their \pkg{recipes} equivalents. At present, \code{tune\_resample()} supports classification and regression workflows, but not survival outcomes.

\subsection{Practical workflow variants}

The same workflow generalizes to the other supported split modes. With \code{batch\_blocked}, all records from the same batch are assigned to the same fold. With \code{study\_loocv}, one study is held out at a time, which is often appropriate for consortium datasets or data aggregated from public resources. With \code{time\_series}, observations are ordered chronologically and training is restricted to records that precede the test block, optionally with additional separation between training and evaluation periods. A \code{combined} mode applies multiple constraints simultaneously (e.g., subject grouping with batch blocking). In all cases, the core principle remains the same: resampling should represent the intended deployment setting rather than a convenient random partition of rows.

\section{Quantifying Leakage Inflation ($\Delta_{\mathrm{LSI}}$)}\label{sec:dlsi}

\pkg{bioLeak} provides \code{delta\_lsi()} to compare a leaky pipeline
with a guarded one when both are fitted on the same repeated resampling design.
The goal is not to decide whether one pipeline is ``better'' in a generic
sense, but to quantify how much apparent performance changes when leakage
controls are introduced. This focus is consistent with the broader
cross-validation literature, which emphasizes that resampling estimates are
themselves random quantities and that their uncertainty is nontrivial to
characterize \citep{bengio2004, bates2024}.

Suppose repeated cross-validation yields $R$ repeat-level summaries. Let
$M_{\mathrm{leaky},r}$ and $M_{\mathrm{guarded},r}$ denote the repeat-level
performance estimates for repeat $r$. For higher-is-better metrics such as AUC
or accuracy, the repeat-level inflation score is
\[
\Delta_r = M_{\mathrm{leaky},r} - M_{\mathrm{guarded},r}, \qquad r = 1, \ldots, R.
\]
Positive values therefore indicate that the leaky pipeline appears more optimistic than the guarded one and, under a properly paired comparison, are consistent with leakage-related performance inflation. For lower-is-better metrics such as RMSE or log-loss, \code{delta\_lsi()} applies a sign flip internally through
the \code{higher\_is\_better} argument so that positive values still denote
inflation.

In the implementation, fold-level metric values are aggregated to a repeat
level using test-set-size weights. Let $\mu_r$ denote the resulting repeat
summary for one pipeline. The vector
$\boldsymbol{\Delta} = (\Delta_1, \ldots, \Delta_R)^\top$ is then summarized in
two ways. The first is the arithmetic mean,
\[
\widehat{\Delta}_{\mathrm{metric}} =
\frac{1}{R} \sum_{r=1}^{R} \Delta_r,
\]
which is easy to interpret directly in the units of the metric. The second is
the robust summary reported as \code{delta\_lsi}. Intuitively, this estimator
behaves like the arithmetic mean for typical repeats but automatically
downweights any repeat whose inflation score is unusually large or small,
providing robustness when one or two folds are unstable for idiosyncratic
reasons. Formally,
\[
\widehat{\Delta}_{\mathrm{LSI}} =
\operatorname*{arg\,min}_{\mu} \sum_{r=1}^{R}
\rho_{1.345} \left( \frac{\Delta_r - \mu}{\widehat{\sigma}} \right),
\]
where $\rho_{1.345}$ is Huber's loss function and
$\widehat{\sigma} = 1.4826 \times \mathrm{MAD}(\boldsymbol{\Delta})$ is a fixed
robust scale estimate. The minimization is solved by iteratively reweighted
least squares (IRLS) \citep{huber1964, huber2009}.

Confidence intervals are computed with a bias-corrected and accelerated (BCa)
bootstrap when the effective number of paired repeats is at least ten. The
package reports a $95\%$ interval for both \code{delta\_lsi} and the
arithmetic mean. Hypothesis testing is based on a sign-flip randomization test.
Under the null hypothesis of no systematic inflation, the sign of each
$\Delta_r$ is exchangeable. The test statistic is $T = R^{-1} \sum_{r=1}^{R}
s_r \Delta_r$, where $s_r \in \{-1, +1\}$ are sign-flip indicators and the
observed value uses $s_r = 1$ for all~$r$. For $R \leq 15$, the package
enumerates all $2^R$ sign combinations exactly. For larger $R$, Monte Carlo
sign-flips are
sampled and the Phipson--Smyth correction is applied so that the reported
$p$-value is never zero \citep{phipson2010}.
When \code{exchangeability = "blocked\_time"}, a block sign-flip procedure
preserves serial autocorrelation under the null. Two further options
(\code{"by\_group"}, \code{"within\_batch"}) are accepted but currently fall
back to the iid procedure with a warning.

The design is only fully inferential when the two compared fits are paired. In
practice this means they must share the same fold structure, not merely the
same number of repeats. If fold membership differs, the package falls back to an
unpaired comparison of repeat-level summaries and suppresses formal inference.
This distinction is important. If the user changes both the predictors and the
resampling design, the resulting performance difference cannot be attributed solely to leakage control.

The returned \code{LeakDeltaLSI} object stores point estimates, intervals,
the sign-flip $p$-value, and repeat-level summaries for both pipelines. The
package groups inferential output into four tiers:
\begin{enumerate}
\item \code{D\_insufficient}: fewer than five effective paired repeats or an
unpaired comparison; point estimates only.
\item \code{C\_signflip}: at least five paired repeats; point estimates and a
sign-flip $p$-value.
\item \code{B\_signflip\_ci}: at least ten paired repeats; point estimates,
$p$-value, and BCa intervals.
\item \code{A\_full\_inference}: at least twenty paired repeats; returns point
estimates, $p$-value, and BCa intervals, and sets \code{inference\_ok = TRUE}
when these are numerically finite. This is the recommended threshold for
trustworthy $\Delta_{\mathrm{LSI}}$ inference.
\end{enumerate}

A typical call compares a leaky and a guarded fit built from the same
\code{LeakSplits} object:

\begin{CodeInput}
inflation <- delta_lsi(
  fit_leaky = fit_leaky,
  fit_guarded = fit_guarded,
  metric = "auc",
  M_boot = 2000,
  M_flip = 10000,
  seed = 42
)

summary(inflation)
\end{CodeInput}

\noindent
The output below is from the case-study comparison described in
Section~\ref{sec:casestudy}, which uses $20$ paired batch-blocked repeats:

\begin{CodeOutput}
=====================================
 bioLeak Delta LSI Summary
=====================================

Metric:            auc
Exchangeability:   iid
Inference tier:    A_full_inference
R_eff:             20  (R_leaky=20, R_guarded=20, paired=TRUE)

Leaky pipeline:    0.791
Guarded pipeline:  0.611

Point estimates (leaky - guarded; positive = leakage inflation):
  delta_metric:  0.180  (raw metric difference)
  95% BCa CI:    [0.172, 0.186]
  delta_lsi:     0.181  (Huber-robust)
  95% BCa CI:    [0.174, 0.187]

Hypothesis test  [H0: no systematic inflation; paired repeat signs exchangeable]:
  Sign-flip p:   0.0001 ***

Inference valid:   YES (tier A)
\end{CodeOutput}

\noindent
In practice, $\Delta_{\mathrm{LSI}}$ is most useful when an analyst wants to
measure the sensitivity of a result to specific leakage controls, such as
subject grouping, batch blocking, or removal of explicit leak-prone features.
Section~\ref{sec:simulation} evaluates the audit framework across a factorial
simulation design, and Section~\ref{sec:casestudy} demonstrates a paired
$\Delta_{\mathrm{LSI}}$ comparison on real data. A small power simulation
(\code{run\_delta\_lsi.R}; $n = 200$, $p = 20$, $50$ seeds, $20$-repeat
$5$-fold CV with a \code{peek\_norm} leakage feature at $\sigma = 0.3$)
yielded $100\%$ rejection at $\alpha = 0.05$ with a mean
$\widehat{\Delta}_{\mathrm{metric}} = 0.197$, confirming that the sign-flip
test reliably detects leakage-induced inflation in this setting.
A complementary null calibration confirmed nominal Type~I error control.
In each of $50$ replications, $R = 20$ independent datasets were drawn from
the same population and both arms received their own exchangeable noise
feature (pure $\mathcal{N}(0,1)$, uncorrelated with the outcome). The
observed rejection rate was $8.0\%$ at $\alpha = 0.05$ (one-sided exact
binomial test against the nominal $5\%$, $H_1$: rate $> 0.05$; $p = 0.24$,
non-significant), with a mean
$\widehat{\Delta}_{\mathrm{metric}}$ of $0.0001$ and balanced sign
proportions across repeats (replication script: \code{run\_delta\_lsi.R},
Part~C).
Under weaker leakage or smaller samples, power would be lower; a
comprehensive power characterization is left for future work.

\section{Simulation Study}\label{sec:simulation}

\subsection{Design}

An important interpretive caveat applies throughout this section. All
permutation-gap analyses use \code{perm\_refit = FALSE}, meaning that the
model is fitted once on the original labels and the permutation procedure
shuffles labels against fixed predictions. This design tests whether the
prediction--label association exceeds chance levels, but it does not
distinguish genuine predictive signal from signal attributable to leakage. As a
result, the clean baseline (\code{none}) can also reject at high rates when
real signal is present ($s > 0$). Leakage-specific detection is therefore most
informative at $s = 0$, where the outcome contains no genuine predictive signal
and any above-chance performance under the leakage conditions must arise from
the injected leakage mechanism.

The manuscript simulations were generated by a dedicated script
(\code{run\_simulation.R}) whose settings differ in several respects from
the package helper \code{simulate\_leakage\_suite()}. These choices were
made to create leakage signals of graded strength for the paper-specific
experiments. In particular, the manuscript script generates peek-norm
leakage as $y + \varepsilon$ with $\varepsilon \sim \mathcal{N}(0, 0.09)$,
implements look-ahead leakage by shifting a continuous biomarker rather
than the binary outcome, and adds AR(1) noise ($\rho = 0.9$) to the latent
linear predictor when $s > 0$. Readers who wish to reproduce the manuscript
results exactly should use the paper script rather than the package helper.

The main simulation follows a full factorial design with five leakage
settings (none, subject overlap, batch confounding, peek normalization, and
look-ahead), four sample sizes ($n \in \{100, 250, 500, 1000\}$), three
feature dimensions ($p \in \{10, 50, 100\}$), three signal levels
($s \in \{0, 0.5, 2.0\}$), and fifty seeds per configuration, yielding
$9{,}000$ simulated datasets in total.

For each dataset, predictors were generated as Gaussian variables and the binary outcome was sampled from a probit link applied to a latent linear predictor. When $s = 0$, the outcome contained no genuine predictive signal. Under the clean condition (\code{none}), this setting provides a null baseline; under the leakage conditions, any above-chance performance is attributable to the injected leakage mechanism rather than to real signal. When $s > 0$, signal entered through the first few predictors, and autoregressive noise was added to the latent process.

Leakage was injected as follows:
\begin{itemize}
    \item \code{subject\_overlap}: a subject-level mean of the binary outcome was added as a predictor;
    \item \code{batch\_confounded}: batch membership was made outcome dependent and the batch-wise mean outcome was added as a predictor;
    \item \code{peek\_norm}: a noisy continuous version of the outcome was added as a globally computed feature; and
    \item \code{lookahead}: a future biomarker value was shifted into the current row. Because the shifted covariate carries outcome information only through the latent signal, this mechanism is structurally null at $s = 0$; it is included primarily to evaluate detection under the $s > 0$ conditions.
\end{itemize}
Preprocessing leakage (Section~\ref{sec:leakage}, Section~2.1) is not directly
simulated as a separate mechanism. The guarded pipeline applies train-fold-only
preprocessing by construction, so the effect of global-vs-fold-wise
preprocessing is absorbed into the package's default behavior rather than
isolated as a controlled factor. A dedicated simulation comparing fold-wise and
global normalization on correlated-feature designs would be a valuable extension
but is beyond the scope of the present study.

All main-simulation analyses used the same five-fold subject-grouped cross-validation design in order to hold the evaluation structure fixed while varying the injected leakage mechanism. The learner was \pkg{glmnet}-penalized logistic regression (elastic net with $\alpha = 0.9$ and cross-validated penalty), with median imputation, z-score normalization, no feature selection, and a permutation-gap audit using $B = 200$ and \code{perm\_refit = FALSE}. The reported summaries include the observed AUC, the permutation-null mean, the
permutation gap, and the permutation $p$-value.

Two supplementary analyses, generated by \code{run\_supplementary.R}, refine this design. The first evaluates the same general leakage mechanisms under four split modes (\code{subject\_grouped}, \code{batch\_blocked}, \code{study\_loocv}, and \code{time\_series}) at a fixed setting $(n = 500, p = 20, s = 1.0)$. The second evaluates the target-scanning components at the same fixed setting.

\subsection{Main simulation summaries}

The main simulation results, generated by \code{run\_simulation.R},
are summarized in Tables~\ref{tab:sim-power} and \ref{tab:sim-s0}. The two
tables separate signal-bearing settings ($s > 0$) from the null setting
($s = 0$) where leakage-specific detection can be interpreted more directly.

\begin{figure}[tbp]
\centering
\includegraphics[width=\textwidth]{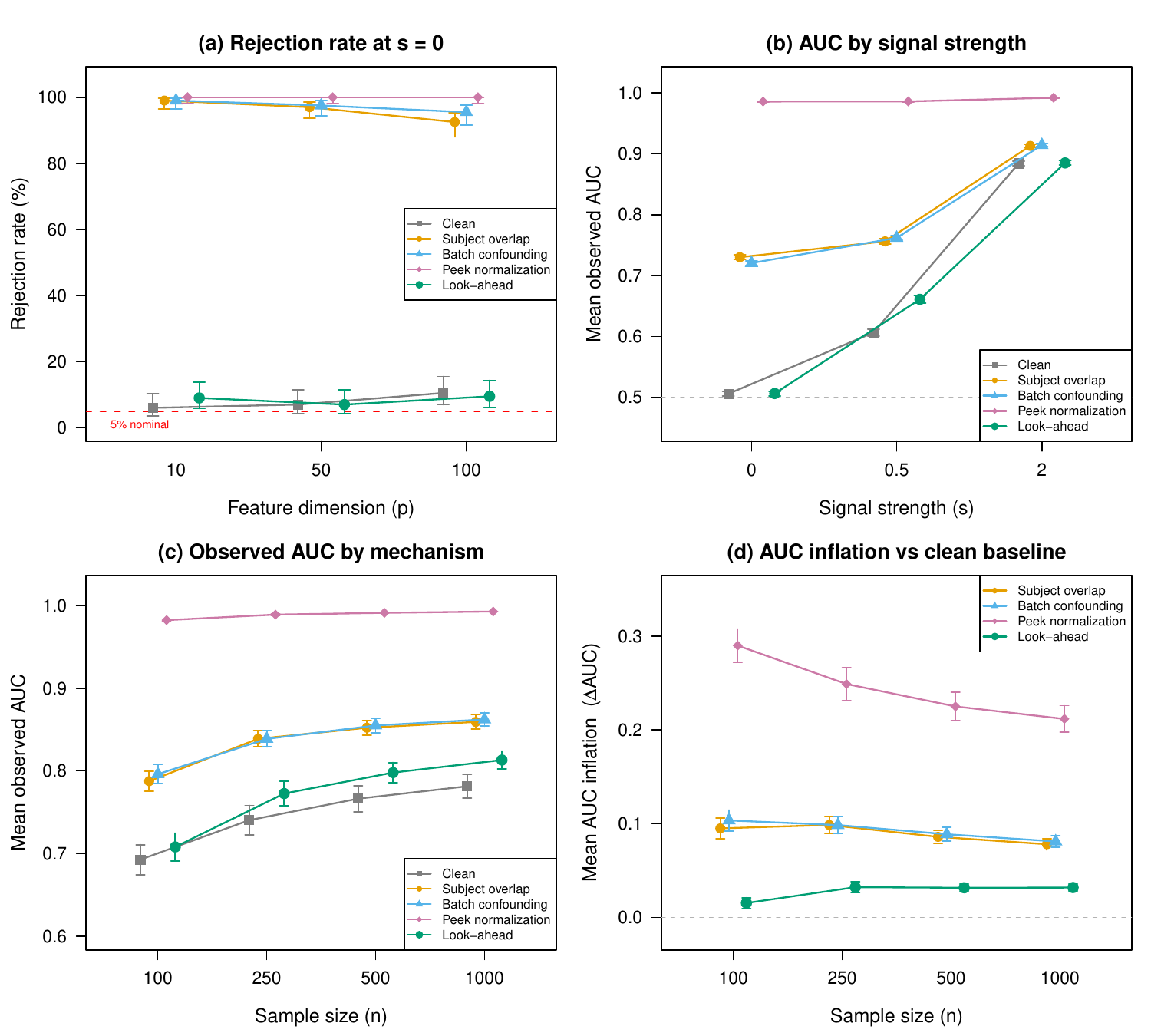}
\caption{Simulation results across leakage mechanisms. (a) Rejection rate
across feature dimensions ($s = 0$). (b) AUC across signal strengths (all $s$).
(c) Observed AUC across sample sizes ($s > 0$). (d) AUC inflation
relative to clean baseline ($s > 0$). Error bars are $95\%$ confidence
intervals: Wilson for the binomial rejection rate in panel (a); Student-$t$
for the AUC panels (b, c, d). Intervals may be too narrow to be
visible at highly precise points. Points are horizontally jittered to
avoid overlap.}
\label{fig:sim-main}
\end{figure}

Figure~\ref{fig:sim-main}(a) shows rejection rates at $s = 0$ (no real signal)
for all mechanisms across feature dimensions. This isolates leakage detection
from genuine predictive signal. Peek normalization is detected at $100\%$
regardless of $p$, while subject overlap and batch confounding are detected at
$92$--$99\%$. In contrast, look-ahead and the clean baseline both remain near
the nominal level ($6$--$10.5\%$), confirming that look-ahead carries no outcome
information at $s = 0$ by design.

\paragraph{Null calibration of the fixed-prediction permutation test.}
For the clean baseline, the overall false rejection rate with
\code{perm\_refit = FALSE} was \textbf{7.8\%} at $\alpha = 0.05$
($8.0\%$ at $n = 100$, $9.3\%$ at $n = 250$, $8.0\%$ at $n = 500$,
$6.0\%$ at $n = 1000$). The inflation scales with feature dimension:
\textbf{6.0\%} at $p = 10$ rising to \textbf{10.5\%} at $p = 100$, and
diminishes with sample size. This mild liberality is a known property of
fixed-prediction permutation tests: the model captures weak noise
correlations during training, producing a small positive gap even when no
true signal exists.
In normal usage this inflation does not arise, because the package
default \code{perm\_refit = "auto"} activates refit-based permutations
whenever \code{fit\_resample()} has stored the original data and learner
(the default \code{store\_refit\_data = TRUE}); see the practical
guidance in Section~\ref{sec:overview}. The simulations in this paper
use explicit \code{perm\_refit = FALSE} because the goal is to
characterize the permutation-gap statistic as a fast association test
under controlled leakage, not to perform full refit-based null
calibration. A refit-based evaluation at the same $B = 200$ would require $200$ full
model fits per task, multiplying the runtime by roughly 100-fold---from
about 75 minutes to over 100 hours on four cores---prohibitive for a
$9{,}000$-task factorial design.

For leakage-injected settings with $s > 0$, the permutation-gap statistic was
usually significant. However, with \code{perm\_refit = FALSE} the
permutation-gap test detects prediction-label association rather than leakage
specifically. Figure~\ref{fig:sim-main}(b) shows how observed AUC varies with signal
strength for each mechanism. At $s = 0$ (no real signal), the clean baseline
sits at chance ($\approx 0.50$), while \code{peek\_norm} already reaches
$0.99$ --- demonstrating that this leakage mechanism creates apparent
performance from nothing. Subject overlap and batch confounding similarly
inflate AUC above chance at $s = 0$ ($\approx 0.72$--$0.73$), whereas
look-ahead is indistinguishable from clean. As $s$ increases, the mechanisms
move toward the clean baseline as genuine signal dominates, though
\code{peek\_norm} retains a small residual inflation from its direct outcome
encoding.

Turning to rejection rates, the clean baseline (\code{none}) also rejects at
high rates when real signal is present: 84.2\% overall at $s > 0$,
reaching $98\%$ at $n = 500$ and $100\%$ at $n = 1000$.
Table~\ref{tab:sim-power} includes this baseline alongside the leakage-injected
runs so that the increment attributable to leakage is visible. Pooled over
feature dimension and the two positive signal levels, the weakest leaky setting
is \code{lookahead} at $n = 100$ ($0.65$), which is close to the clean
baseline at the same $n$ ($0.57$). Subject overlap, batch confounding, and
\code{peek\_norm} exceed the clean baseline by a larger margin at $n = 100$
and reach $1.00$ by $n = 250$.
Each cell in Table~\ref{tab:sim-power} pools $300$ runs
($3$ feature dimensions $\times$ $2$ signal levels $\times$ $50$ seeds);
standard errors are binomial and shown in parentheses. Cells at $1.00$
have zero Wald standard error; the Wilson $95\%$ lower bound is $0.99$
for $300/300$.

\begin{table}[tbp]
\centering
\caption{Rejection rates (permutation-gap $p < 0.05$) at $s > 0$, pooled
over $p$ and $s$.}
\label{tab:sim-power}
\begin{tabular}{|l|c|c|c|c|}
\hline
\textbf{Mechanism} & \textbf{$n=100$} & \textbf{$n=250$} & \textbf{$n=500$} & \textbf{$n=1000$} \\
\hline
None (clean baseline) & 0.57 (.03) & 0.82 (.02) & 0.98 (.01) & 1.00 \\
\hline
Subject overlap & 0.93 (.01) & 1.00 & 1.00 & 1.00 \\
\hline
Batch confounding & 0.95 (.01) & 1.00 & 1.00 & 1.00 \\
\hline
Peek normalization & 1.00 & 1.00 & 1.00 & 1.00 \\
\hline
Look-ahead & 0.65 (.03) & 0.95 (.01) & 1.00 & 1.00 \\
\hline
\end{tabular}
\end{table}

\noindent

These values should be interpreted carefully. With
\code{perm\_refit = FALSE}, the permutation-gap test detects non-random
predictive signal rather than leakage in isolation, so the high baseline
rejection rate at $s > 0$ means that absolute rejection rates cannot be
read as leakage-detection power. A more informative view comes from the
$s = 0$ condition, where no real signal exists and the only source of
above-chance performance is the injected leakage. Table~\ref{tab:sim-s0} shows the
leakage-specific detection rates under this design. Three of the four
mechanisms are detected reliably by $n = 250$; look-ahead is not detectable
at $s = 0$ because the shifted biomarker carries no outcome information
when the latent signal is absent.
Each cell pools $150$ runs ($3$ feature dimensions $\times$ $50$ seeds);
standard errors and Wilson bounds follow the same conventions as
Table~\ref{tab:sim-power} (Wilson $95\%$ lower bound $0.98$ for $150/150$).

\begin{table}[tbp]
\centering
\caption{Leakage-specific detection at $s = 0$ (no real signal), pooled
over $p$.}
\label{tab:sim-s0}
\begin{tabular}{|l|c|c|c|c|}
\hline
\textbf{Mechanism} & \textbf{$n=100$} & \textbf{$n=250$} & \textbf{$n=500$} & \textbf{$n=1000$} \\
\hline
None (clean baseline) & 0.08 (.02) & 0.09 (.02) & 0.08 (.02) & 0.06 (.02) \\
\hline
Subject overlap & 0.85 (.03) & 1.00 & 1.00 & 1.00 \\
\hline
Batch confounding & 0.89 (.03) & 1.00 & 1.00 & 1.00 \\
\hline
Peek normalization & 1.00 & 1.00 & 1.00 & 1.00 \\
\hline
Look-ahead & 0.09 (.02) & 0.09 (.02) & 0.09 (.02) & 0.06 (.02) \\
\hline
\end{tabular}
\end{table}

\noindent
The $6$--$9\%$ clean-baseline rates in Table~\ref{tab:sim-s0} reflect the
same mild liberality discussed in the null calibration paragraph above; see
Section~\ref{sec:workflow} for the refit-based alternative.

This two-table presentation separates the permutation test's role as a general
performance check ($s > 0$) from its role as a leakage detector ($s = 0$). The
package combines the permutation module with targeted diagnostics for
confounding, duplicates, and target leakage precisely because no single
component is leakage-specific in all settings.

Figure~\ref{fig:sim-main}(c) shows the observed AUC for each mechanism across
sample sizes (averaged over $s > 0$ conditions). Peek normalization inflates
AUC to near~$1.0$ regardless of $n$, while subject overlap and batch
confounding produce moderate inflation that narrows as $n$ grows and genuine
signal dominates. Look-ahead leakage sits only slightly above the clean
baseline, consistent with the weak temporal correlation in this simulation
design. To quantify inflation directly, panel~(d) compares each leakage
mechanism with the clean baseline at matched $(n, s)$ values. Averaged over the
simulation configurations with $s > 0$, the mean observed AUC in clean settings was
$0.75$. Relative to that clean baseline, the mean inflation in observed AUC was
$0.089$ for \code{subject\_overlap}, $0.093$ for
\code{batch\_confounded}, $0.244$ for \code{peek\_norm}, and $0.028$ for
\code{lookahead}. In these simulations, explicit global peek-type leakage
produces the largest inflation, whereas the look-ahead construction yields the
smallest average AUC increase. This difference is attributable to the
particular manuscript design in \code{run\_simulation.R}; it should not
be read as a universal ranking of leakage severity.

\subsection{Supplementary split-mode and target-scan analyses}

Table~\ref{tab:sim-modes} summarizes the supplementary split-mode analysis
from \code{run\_supplementary.R} in the replication materials. As with the main
simulation, the split-mode experiment was run at both $s = 0$ (leakage-specific)
and $s = 1.0$ (signal present). At $s = 1.0$, the clean baseline rejects in all
$50/50$ runs for every split mode, so the $s = 1.0$ results cannot isolate
leakage. Table~\ref{tab:sim-modes} therefore reports the $s = 0$ condition,
where the only source of above-chance performance is the injected leakage.
Across all four split modes, the \code{subject\_overlap},
\code{batch\_confounded}, and \code{peek\_norm} mechanisms are detected with
high power, with the notable exception of \code{batch\_confounded} under
\code{batch\_blocked} mode ($0.02$), where the blocking variable aligns
with the confounding structure. Look-ahead leakage is largely undetectable at
$s = 0$ (rejection rates $0.02$--$0.04$ across modes), which is
expected: this mechanism leaks future values of a continuous biomarker that
carries signal only when real signal is present.
Each cell is based on $50$ seeds; standard errors in parentheses.
Cells at $1.00$ have Wilson $95\%$ lower bound $0.93$ for $50/50$.

\begin{table}[tbp]
\centering
\caption{Supplementary split-mode analysis: rejection rates
(permutation-gap $p < 0.05$) at $(n = 500, p = 20, s = 0)$.}
\label{tab:sim-modes}
\small
\begin{tabular}{|l|c|c|c|c|c|}
\hline
\textbf{Mode} & \textbf{None} & \textbf{Subj.\ overlap} & \textbf{Batch conf.} & \textbf{Peek norm.} & \textbf{Look-ahead} \\
\hline
Subj.\ grouped & 0.06 (.03) & 1.00 & 0.98 (.02) & 1.00 & 0.04 (.03) \\
\hline
Batch blocked  & 0.04 (.03) & 1.00 & 0.02 (.02) & 1.00 & 0.02 (.02) \\
\hline
Study LOOCV    & 0.06 (.03) & 1.00 & 0.98 (.02) & 1.00 & 0.04 (.03) \\
\hline
Time series    & 0.02 (.02) & 1.00 & 0.98 (.02) & 1.00 & 0.02 (.02) \\
\hline
\end{tabular}
\end{table}

The reduced detection of \code{batch\_confounded} leakage under
\code{batch\_blocked} mode is an expected interaction: when the test fold
contains samples from a single batch, confounding between batch and outcome is
inherently difficult to detect via permutation because the batch structure is
held constant. This illustrates a key design lesson: the permutation-gap test's
sensitivity depends on the interplay between the leakage mechanism and the
resampling geometry. The exercise reinforces the core design principle of
\pkg{bioLeak}: the choice of split design is part of the model
specification, and combining permutation tests with targeted diagnostics
(such as the target scan) provides more robust detection.

\begin{figure}[tbp]
\centering
\includegraphics[width=0.65\textwidth]{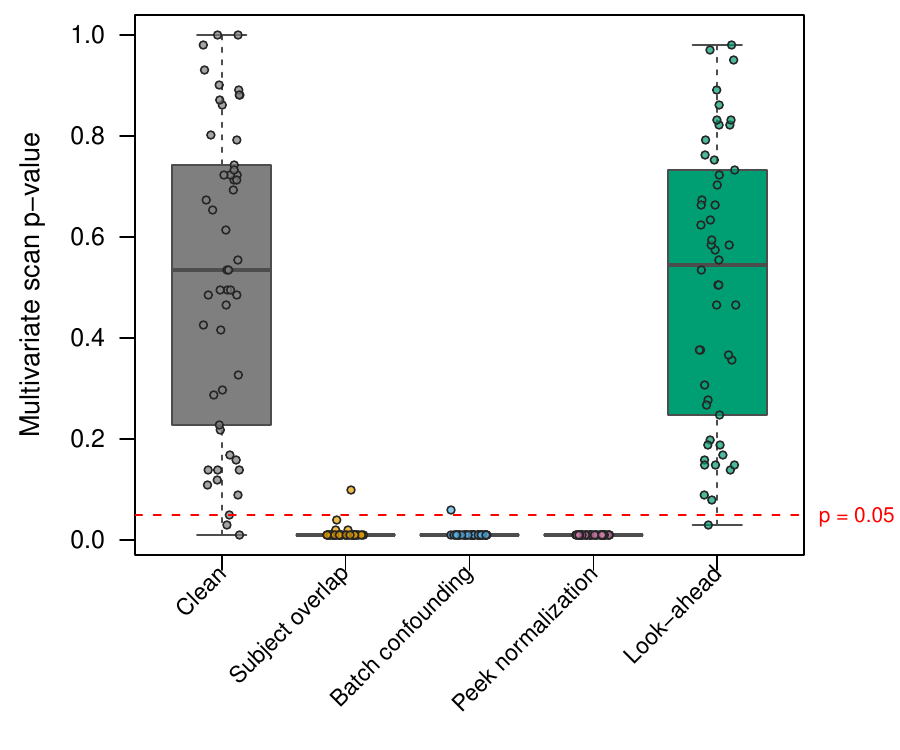}
\caption{Multivariate target-scan $p$-values at $s = 0$.}
\label{fig:sim-target}
\end{figure}

The supplementary target-scan analysis, also from
\code{run\_supplementary.R}, provides a more
granular picture. At $s = 0$ (no real signal), the univariate target scan
flagged the injected leakage feature in all $50/50$ \code{peek\_norm} runs
and in none of the \code{subject\_overlap}, \code{batch\_confounded},
\code{lookahead}, or clean runs at the threshold used in the stored script
(\code{target\_threshold = 0.9}). Averaged over all leaky conditions, this
corresponds to a 25\% success rate because only one of the four
injected mechanisms satisfied the conservative univariate threshold. The
multivariate scan at $s = 0$ rejected in 74\% of leaky runs, with
a 6\% false-positive rate in the clean baseline
(Figure~\ref{fig:sim-target})---consistent with
the nominal $\alpha = 0.05$ level. This contrasts with the $s = 1.0$
evaluation, where the multivariate scan rejects in all conditions (including
the clean control) because the real signal features dominate the multivariate
scan's test statistic, which is the cross-validated predictive score of a
GLM fitted on principal components of \code{X\_ref}. The $s = 0$ results confirm that the multivariate scan has genuine
leakage-detection power when confounding with real signal is removed.
Nonetheless, the univariate and multivariate scans should be interpreted
jointly, and with knowledge of the reference matrix supplied to
\code{X\_ref}.

\section{Case Study}\label{sec:casestudy}

\subsection{Data and workflow}

The case study is based on \code{run\_casestudy.R} in the replication
materials. The script draws data
from the \pkg{curatedOvarianData} Bioconductor package \citep{ganzfried2013} and constructs a
binary three-year overall survival endpoint from studies containing both
\code{days\_to\_death} and \code{vital\_status}. After intersecting feature
sets across eligible studies and retaining the $2{,}000$ most variable common
genes, the final analysis dataset contains $N = 2{,}015$ samples from $S = 14$
studies and $G = 2{,}000$ genes. Study sizes range from 50 to 376 samples, with a
median of 125. Gene-expression prediction has long served as a
test case for both high-dimensional classifiers and evaluation pitfalls in
biomedical machine learning \citep{golub1999, simon2003, tarca2007}.

Four workflows were evaluated to permit a clean decomposition of the
performance gap. The \emph{guarded} workflow used \code{study\_loocv}, median
imputation, z-score normalization, t-test feature selection (FS) with the top 100
genes per fold, and a \pkg{glmnet} learner. The \emph{leaky} comparator used
standard five-fold row-wise evaluation implemented via a
\code{subject\_grouped} split on a dummy row identifier, the same learner,
and three deliberately leak-prone features: a per-study outcome mean, the
binary outcome with additive Gaussian noise ($\sigma = 1.0$), and a globally
standardized first principal component.

Because the guarded and leaky pipelines differ in three ways simultaneously
(feature selection, split design, and leakage features), attributing the
AUC gap to any single factor requires intermediate arms. Two sensitivity
pipelines bridge the extremes:
\begin{enumerate}
\item \textbf{Guarded (no FS)}: same study-LOOCV splits and clean data as the
  guarded pipeline, but no feature selection---isolates the feature-selection
  effect.
\item \textbf{Naive (no leaks)}: same row-wise 5-fold split as the leaky pipeline,
  but with clean data and no feature selection---isolates the split-design
  effect from the leakage-feature effect.
\end{enumerate}

All four workflows were audited with a permutation-gap test
($B = 500$), study-association testing, target scans, and duplicate detection.

\subsection{Results}

Table~\ref{tab:case-study} summarizes the case-study output. The guarded
workflow achieved an AUC of $0.610$ with fold-to-fold standard deviation
$0.063$, whereas the leaky comparator achieved an AUC of $0.810$ with standard
deviation $0.024$. The total AUC gap is therefore $0.200$. The four-arm design
decomposes this gap into three additive components:
\begin{enumerate}
\item \textbf{Feature-selection effect} ($+0.072$): guarded $\to$ guarded (no FS), both
  with study-LOOCV and clean data. Removing per-fold t-test feature selection
  raises the AUC from $0.610$ to $0.682$. This increase reflects the fact that
  \pkg{glmnet}'s built-in $L_1$ regularization selects relevant features more
  flexibly across all $2{,}000$ genes than a rigid top-100 univariate filter.
\item \textbf{Split-design effect} ($+0.035$): guarded (no FS) $\to$ naive
  (no leaks), both with clean data and no feature selection. Switching from
  study-LOOCV to row-wise 5-fold CV raises the AUC from $0.682$ to $0.717$.
  The increase represents optimistic bias: row-wise splitting allows samples
  from the same study to appear in both training and test sets, so
  study-specific expression patterns inflate the estimate.
\item \textbf{Leakage-feature effect} ($+0.093$): naive (no leaks) $\to$ leaky,
  both with row-wise 5-fold CV. Adding three leak-prone features raises the
  AUC from $0.717$ to $0.810$. Even with substantial noise ($\sigma = 1.0$),
  the outcome-derived feature provides the model with direct label information,
  demonstrating that noisy proxies of the target can cause substantial
  inflation.
\end{enumerate}
The leakage features account for $47\%$ of the raw AUC gap, feature selection
for $36\%$, and the split-design change for $17\%$. Because the decomposition
is sequential, these percentages are path-dependent: a different ordering of
the three factors would redistribute the attributions. The chosen order
(feature selection $\to$ split design $\to$ leakage features) moves from the
most methodological choice to the most explicit data contamination, which we
consider the most interpretable progression.  The permutation gap is
$0.101$ (guarded), $0.173$ (guarded, no FS), $0.208$ (naive, no leaks), and
$0.301$ (leaky). All four permutation null distributions center near $0.5$
by construction (global label shuffle), and all $p$-values equal $0.002$
(the floor at $B = 500$).

\begin{table}[tbp]
\centering
\caption{Case-study summary. The four columns form a decomposition chain:
each successive pipeline changes exactly one factor (feature selection, split
design, or leakage features), isolating its contribution to the AUC gap.
Study-fold $V$ is Cram\'er's $V$ between outer fold assignment and study
identity; the multivariate scan $p$ is from the target-scan audit.}
\label{tab:case-study}
\small
\begin{tabular}{|l|c|c|c|c|}
\hline
 & \textbf{Guarded} & \textbf{Guarded (no FS)} & \textbf{Naive (no leaks)} & \textbf{Leaky} \\
\hline
Split design & Study LOOCV & Study LOOCV & 5-fold CV & 5-fold CV \\
Feature selection & t-test top 100 & None & None & None \\
Leak features & No & No & No & Yes \\
AUC & 0.610 & 0.682 & 0.717 & 0.810 \\
SD & 0.063 & 0.079 & 0.025 & 0.024 \\
Perm.\ gap & 0.101 & 0.173 & 0.208 & 0.301 \\
Perm.\ $p$ & 0.002 & 0.002 & 0.002 & 0.002 \\
Study-fold $V$ & 1.000\textsuperscript{$\dagger$} & 1.000\textsuperscript{$\dagger$} & 0.079\textsuperscript{$\ddagger$} & 0.079 \\
Multivar.\ scan $p$ & 0.2673 & ---\textsuperscript{$\S$} & ---\textsuperscript{$\S$} & 0.0099 \\
\hline
\multicolumn{5}{l}{\footnotesize\textsuperscript{$\dagger$}$V = 1$ is expected: each study-LOOCV fold is defined by study, so fold} \\
\multicolumn{5}{l}{\footnotesize assignment and study identity are perfectly associated by design.} \\
\multicolumn{5}{l}{\footnotesize\textsuperscript{$\ddagger$}Same split as the leaky pipeline (dummy-subject 5-fold).} \\
\multicolumn{5}{l}{\footnotesize\textsuperscript{$\S$}Not computed for the sensitivity pipelines (audit limited to perm.\ gap).} \\
\end{tabular}
\end{table}

The study-fold association results deserve explicit interpretation. In the
guarded workflow, Cram\'er's $V$ is exactly $1.000$ because each outer fold is
defined by study. This is not evidence of an implementation error; it is a
property of the intended external-validation design. In the leaky comparator,
the association is weak ($V = 0.079$) because the row-wise split ignores study
boundaries. This contrast illustrates why association tables must be
interpreted against the chosen resampling policy rather than in isolation.

\begin{figure}[tbp]
\centering
\includegraphics[width=\textwidth]{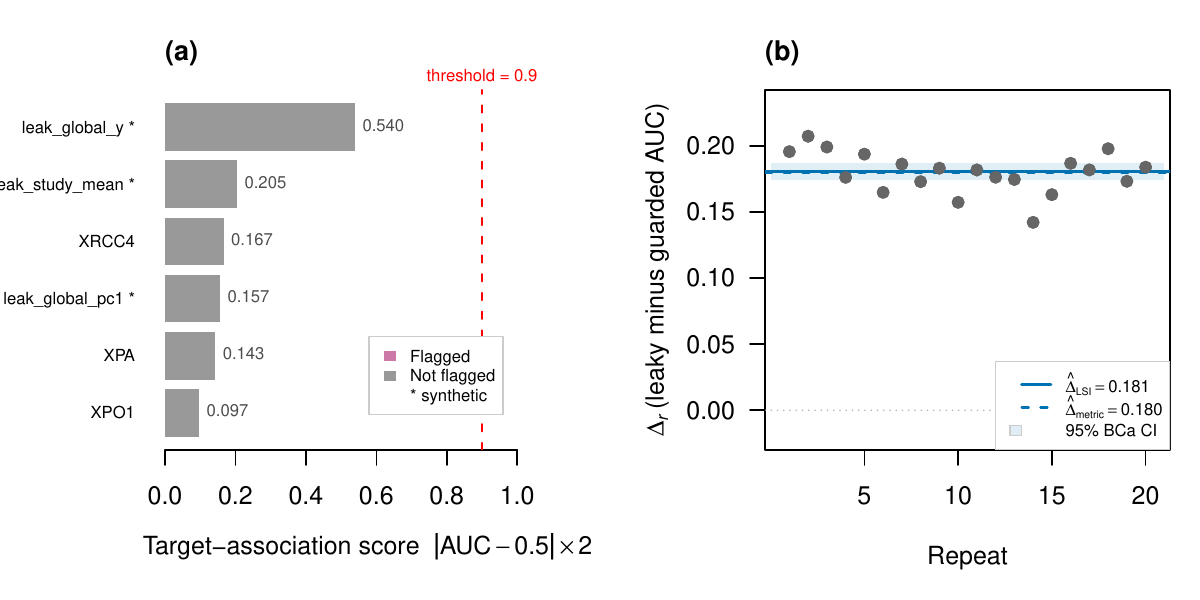}
\caption{Case-study summaries. (a) Top univariate target-association scores
(rescaled AUC, $|\text{AUC} - 0.5| \times 2$) for the leaky pipeline feature
matrix; asterisks mark injected synthetic leakage features. (b) Repeat-level
inflation scores $\Delta_r$ (leaky minus guarded AUC) from
$\Delta_{\mathrm{LSI}}$ with 20 paired repeats; the solid line is the robust
Huber estimate $\widehat{\Delta}_{\mathrm{LSI}}$, the dashed line is the
arithmetic mean, and the shaded band is the $95\%$ BCa bootstrap interval.}
\label{fig:case-study}
\end{figure}

The target-scan results illustrate the complementary value of multiple audit
components (Figure~\ref{fig:case-study}(a)). In the guarded workflow,
no feature exceeded the conservative univariate threshold. The top
association scores correspond to genes \code{XRCC4}, \code{XPA}, and
\code{XPO1}, with scores $0.167$, $0.143$, and $0.097$, respectively. In the
leaky comparator, the noisy outcome feature \code{leak\_global\_y} had the
highest score ($0.540$) but did \emph{not} exceed the $0.9$ flagging
threshold. The weaker proxy features \code{leak\_study\_mean} and
\code{leak\_global\_pc1} had scores $0.205$ and $0.157$. Despite the absence
of univariate flags, the multivariate target scan distinguished the two
workflows: the guarded analysis yielded $p = 0.2673$ (not
significant at $\alpha = 0.05$), whereas the leaky comparator yielded
$0.0099$. This pattern---a moderate univariate signal that escapes
individual-feature flagging but is captured by the joint test---illustrates why
the audit combines both scan types.

Duplicate detection returned a large number of unique near-duplicate sample
pairs in both workflows: $21{,}435$ pairs in the
guarded workflow and $33{,}885$ in the leaky workflow. Pairs were identified
by cosine similarity on the reference expression matrix (threshold $0.995$),
canonicalized, and filtered to those that crossed training and test partitions
in at least one fold (each pair counted once regardless of how many folds it
spans). These likely correspond to technical replicates or near-identical
expression profiles from the same patient across studies.
The count is sensitive to the similarity threshold (here $0.995$) and
should be interpreted relative to the expected technical variability in
the assay.
This result is important for interpretation: leakage-aware
study holdout alone does not guarantee independence when public data resources
contain replicated or near-replicated samples across studies. In this case
study, the study-LOOCV design already keeps same-study replicates on the same
side of each fold, so the duplicates do not inflate the guarded AUC; however,
in the naive 5-fold arms, near-identical samples from different studies can
appear in both training and test sets, contributing to the observed performance
gap. The duplicate screen therefore contributes information that is not
captured by the resampling policy itself.

To quantify the AUC gap formally, we applied $\Delta_{\mathrm{LSI}}$ to the
paired fits on the same batch-blocked folds with 20 repeats
(\code{run\_delta\_lsi.R}; Figure~\ref{fig:case-study}(b)).
The arithmetic mean inflation was $\widehat{\Delta}_{\mathrm{metric}}
= 0.180$ AUC, and the robust Huber estimate was
$\widehat{\Delta}_{\mathrm{LSI}} = 0.181$ AUC, with a $95\%$ BCa
bootstrap interval of $[0.174,\; 0.187]$. As
Figure~\ref{fig:case-study}(b) shows, all $20$ repeat-level deltas are
positive and tightly clustered between $0.14$ and $0.21$, with no outlying
repeats---consistent with the near-identical robust and arithmetic estimates.
The sign-flip randomization test yielded $p < 0.001$, supporting the
conclusion that the leaky pipeline exhibits systematic performance inflation
relative to the guarded one. This pairing is
important: because the leaky and guarded pipelines share the same fold
assignments, the repeat-level differences isolate the effect of leakage
controls from the variance of fold sampling. The full \code{summary()}
output for this \code{LeakDeltaLSI} object is shown in the code block in
Section~\ref{sec:dlsi}. Note that the $\Delta_\mathrm{LSI}$ captures
the combined effect of all three factors; the four-arm decomposition in
Table~\ref{tab:case-study} is needed to separate their individual
contributions.

Overall, the case study demonstrates the intended use of the package. A guarded
workflow produces a more conservative cross-study estimate, while the audit
components identify explicit target leakage and extensive near-duplicate overlap
in the leaky comparator. No single audit component tells the full story:
the permutation gap quantifies inflation, the multivariate target scan detects
leakage features that escape univariate screening, and the duplicate detector
reveals sample-level non-independence that persists even under study-blocked
evaluation. The four-arm decomposition shows that nearly half (47\%) of the
AUC gap is attributable to explicit leakage features, while the remainder
reflects methodological choices (feature selection and split design) that are
not leakage per se but still affect the conservativeness of the estimate.
The exercise also illustrates that some audit components, such as study-fold
association, depend on the evaluation design and must be interpreted in context.

\section{Implementation}\label{sec:implementation}

\subsection{Object structure and pipeline components}

The package uses S4 classes (\code{LeakSplits}, \code{LeakFit},
\code{LeakAudit}, \code{LeakDeltaLSI}; see Section~\ref{sec:overview})
to maintain explicit state across the workflow. Because fits and audits are
self-contained objects, auditing can be decoupled from fitting---useful
when model fitting is computationally expensive.

Guarded preprocessing estimates all transformation parameters on the training
fold and stores a transform function for applying them to the test fold.
Supported operations include imputation (median, kNN, random-forest-based),
normalization (z-score, robust scaling), variance and IQR filtering, and
feature selection (t-test filter, lasso-based selection, or
principal-component projection). Character and factor levels are carried
from training to test fold automatically.

Split construction encodes design decisions explicitly. Each mode uses the
appropriate grouping or ordering column; \code{time\_series} additionally
supports prediction-horizon, purge, and embargo windows. A compact storage
option reduces memory by storing a per-row fold vector rather than explicit
index lists.

\subsection{Reproducibility features}

Several implementation details support reproducible analysis. Split objects
store seeds and a content hash. Fit objects store the original data and
learner configuration by default (\code{store\_refit\_data = TRUE}),
enabling refit-based permutation tests without manual setup. Audit objects
store both the parameters used and, optionally, the permutation draws
themselves. This design is consistent with the principle that
computational claims should remain inspectable at the level of code, data
products, and derived summaries \citep{gentleman2007, peng2011, sandve2013}.

\subsection{Dependency structure}

The core package has few hard dependencies. Optional imports extend
functionality: \pkg{parsnip} and \pkg{recipes} for \pkg{tidymodels}
interoperability, \pkg{glmnet} and \pkg{ranger} for common learners,
\pkg{survival} for time-to-event metrics, and \pkg{rmarkdown} for HTML
report rendering. The package remains usable for basic workflows even when
some optional tools are not installed.

\subsection{Relationship to existing software}

\pkg{bioLeak} is not a general replacement for \proglang{R} machine-learning
frameworks such as \pkg{caret}, \pkg{tidymodels}, or \pkg{mlr3}. Its
role is narrower and complementary. Those frameworks provide broad model
interfaces, preprocessing tools, and resampling utilities. \pkg{bioLeak}
focuses specifically on leakage-aware split construction, guarded preprocessing,
post hoc leakage diagnostics, and structured reporting. While \pkg{mlr3pipelines} \citep{lang2019, bischl2024} can encapsulate preprocessing
within resampling, \pkg{bioLeak} adds a diagnostic layer---permutation-gap
testing, target scans, duplicate detection, and structured reporting---that
is not provided by general-purpose pipeline frameworks. The \pkg{fastml}
package \citep{korkmaz2026} provides guarded resampling workflows for
automated machine learning, re-estimating preprocessing within each resample
to prevent leakage-induced performance inflation; \pkg{bioLeak} is
complementary, focusing on the evaluation design and post hoc audit layer
rather than the model-search and training loop. The package can be used
independently, but it is designed to coexist with existing modeling tools
rather than to supersede them \citep{kuhn2008, kuhn2020, lang2019, bischl2024}.

\section{Discussion}\label{sec:discussion}

\subsection{Design perspective}

\pkg{bioLeak} addresses a practical gap in biomedical machine-learning
workflows by pairing leakage-aware prevention mechanisms with post hoc
diagnostic tools. The package design reflects a simple principle: performance
estimation and leakage assessment should be treated as parts of one workflow,
not as separate stages owned by different pieces of software. This integration
is useful in biomedical settings because the causes of leakage are often spread
across data preparation, resampling, and study design.

The package has several strengths. First, it makes the resampling geometry
explicit. Users must choose whether the relevant independence unit is subject,
batch, study, or time. Second, train-fold-only preprocessing is built into the
central fitting interface. Third, the auditing framework creates a record of
what was checked and how. Fourth, $\Delta_{\mathrm{LSI}}$ offers a direct way to
quantify how much reported performance changes when leakage controls are
introduced. Taken together, these components support a style of model
assessment that is more transparent than a single cross-validated AUC or
accuracy estimate. In that sense, the package makes one practical response to
the reproducibility concerns emphasized by \citet{kapoor2023}: leakage
checks become part of the analysis record rather than an informal afterthought.

\subsection{Limitations}

Several limitations should be stated clearly. First, the package does not prove
the absence of leakage. Audit components only test the evidence available in
the stored predictions, metadata, and optional reference matrix. If a relevant
grouping variable is omitted, no audit procedure can reconstruct it. Second,
unsupervised learning workflows are not currently supported; the package is
organized around supervised outcomes and supervised or outcome-linked audit
procedures. Third, complex graph-based dependency modeling is not yet
implemented. Current split construction is based on subject, batch, study,
time, and related grouped constraints rather than on arbitrary dependency
graphs.

The modeling interface supports binary classification, multiclass
classification, regression, and survival analysis, but function-level
coverage varies by task type. Table~\ref{tab:task-coverage} summarizes the
current state. The audit diagnostics and simulation study presented here focus
on binary classification. The permutation-gap test and $\Delta_{\mathrm{LSI}}$
extend naturally to regression and survival metrics, but have not been
evaluated empirically beyond classification in this manuscript.

\begin{table}[tbp]
\centering
\caption{Function-level task-type coverage. \checkmark\ = supported,
(\checkmark) = supported but not empirically evaluated in this paper,
--- = not yet implemented.}
\label{tab:task-coverage}
\small
\begin{tabular}{|l|c|c|c|c|}
\hline
\textbf{Function} & \textbf{Binary} & \textbf{Multiclass} & \textbf{Regression} & \textbf{Survival} \\
\hline
\code{make\_split\_plan} & \checkmark & \checkmark & \checkmark & \checkmark \\
\code{fit\_resample} & \checkmark & \checkmark & \checkmark & \checkmark \\
\code{tune\_resample} & \checkmark & \checkmark & \checkmark & --- \\
\code{audit\_leakage} & \checkmark & (\checkmark) & (\checkmark) & (\checkmark) \\
\code{delta\_lsi} & \checkmark & (\checkmark) & (\checkmark) & (\checkmark) \\
Target scan & \checkmark & (\checkmark) & (\checkmark) & --- \\
\code{audit\_report} & \checkmark & (\checkmark) & (\checkmark) & (\checkmark) \\
\hline
\end{tabular}
\end{table}

Additional limitations are more specific to the current diagnostics. The
univariate target scan is conservative by design and can miss indirect or
multivariate proxies, while the stored supplementary results show that the
multivariate scan may respond to general predictive signal rather than to
leakage alone. Duplicate detection depends on the supplied feature space and
the chosen similarity threshold. Association tables can reflect the intended
split policy rather than a defect, as the case study illustrates for study
holdout. The \code{delta\_lsi()} exchangeability options
\code{by\_group} and \code{within\_batch} fall back to the iid engine
with a warning (Section~\ref{sec:dlsi}); specialized randomization schemes
for these structures are not yet implemented.

\subsection{Future extensions}

These restrictions are implementation choices rather than conceptual
impossibilities, and they define concrete directions for further development.
Future work could expand the leakage taxonomy to include external-knowledge
leakage, improve multivariate target-leakage calibration, extend nested tuning
to survival models, and add more formal summaries for duplicate-risk severity.
Another useful extension would be tighter integration with manuscript
generation, so that audit reports, scalar summaries, and figure artifacts can be
linked automatically in literate-analysis pipelines.

\section{Summary}\label{sec:summary}

\pkg{bioLeak} provides an integrated \proglang{R} workflow for preventing
and diagnosing data leakage in biomedical machine learning. The package pairs
leakage-aware resampling and guarded preprocessing with post hoc permutation
tests, batch-association diagnostics, target-leakage scans, and duplicate
detection. The $\Delta_{\mathrm{LSI}}$ framework offers formal inference for
comparing leaky and guarded pipelines. The simulation study shows that the
permutation-gap test reliably detects several leakage mechanisms at moderate
sample sizes, while the case study demonstrates how audit components interact
on real multi-study transcriptomic data. Together, these tools make leakage
assessment a documented, reproducible part of the modeling record.

\section{Availability}\label{sec:availability}

\pkg{bioLeak} is available on CRAN, with a long-form tutorial as the
package vignette (\code{bioLeak-intro}). The development source repository
is hosted at \url{https://github.com/selcukorkmaz/bioLeak}. The package is
released under the MIT license.

\subsection*{Replication materials}

All results reported in this manuscript are drawn from stored,
seed-deterministic outputs. The main simulation comprises $9{,}000$ tasks
($180$ configurations $\times$ $50$ seeds), with a total runtime of about
$75$~minutes on four cores. Because routine re-execution during review is
impractical, results are distributed as stored \code{.rds} files in the
replication materials.

Four self-contained generating scripts each set a deterministic seed and
write per-task checkpoints: \code{run\_simulation.R} runs the main
factorial simulation of Section~\ref{sec:simulation},
\code{run\_supplementary.R} runs the splitting-mode robustness and
target-scan analyses also reported in that section,
\code{run\_casestudy.R} runs the \pkg{curatedOvarianData} case study of
Section~\ref{sec:casestudy}, and \code{run\_delta\_lsi.R} runs the
$\Delta_{\mathrm{LSI}}$ case-study and power analyses. A compilation script
(\code{compile\_results.R}) derives every number in the manuscript from
these stored objects, so a reader can audit the full chain from raw outputs
to reported values. The checkpoint mechanism allows partial reruns of
individual scripts. Replication materials are distributed as
supplementary files with this manuscript and through the GitHub repository,
separately from the CRAN source tarball.

\section*{Computational details}

The results in this paper were obtained using
\proglang{R}~4.5.2 \citep{R}. \proglang{R} itself
and all packages used are available from the Comprehensive
\proglang{R} Archive Network (CRAN) at
\url{https://CRAN.R-project.org/} and Bioconductor at
\url{https://bioconductor.org/}.

The package requires \proglang{R}~$\geq$~4.3.
The principal package is \pkg{bioLeak}~0.3.6. Key dependencies include
\pkg{parsnip} for model specification,
\pkg{recipes} for preprocessing pipelines,
\pkg{glmnet} for regularized regression,
\pkg{pROC} for AUC computation,
and \pkg{future} and \pkg{future.apply} for parallel
execution. Exact version numbers for the analysis environment used in this
paper are recorded in the \code{sessionInfo()} output included in the
replication materials (\code{session\_info.txt}).

\bibliographystyle{plainnat}
\bibliography{refs}

@article{ganzfried2013,
  author    = {Ganzfried, Benjamin Frederick and Riester, Markus and Haibe-Kains, Benjamin and Risch, Thomas and Tyekucheva, Svitlana and Jazic, Ina and Wang, Xin Victoria and Ahmadifar, Mahnaz and Birrer, Michael and Parmigiani, Giovanni and Huttenhower, Curtis and Waldron, Levi},
  title     = {{curatedOvarianData}: Clinically Annotated Data for the Ovarian Cancer Transcriptome},
  journal   = {Database},
  year      = {2013},
  volume    = {2013},
  doi       = {10.1093/database/bat013}
}

@article{ambroise2002,
  author    = {Ambroise, Christophe and McLachlan, Geoffrey J.},
  title     = {Selection Bias in Gene Extraction on the Basis of Microarray Gene-Expression Data},
  journal   = {Proceedings of the National Academy of Sciences},
  year      = {2002},
  volume    = {99},
  number    = {10},
  pages     = {6562--6566},
  doi       = {10.1073/pnas.102102699}
}

@article{arlot2010,
  author    = {Arlot, Sylvain and Celisse, Alain},
  title     = {A Survey of Cross-Validation Procedures for Model Selection},
  journal   = {Statistics Surveys},
  year      = {2010},
  volume    = {4},
  pages     = {40--79},
  doi       = {10.1214/09-SS054}
}

@article{bates2024,
  author    = {Bates, Stephen and Hastie, Trevor and Tibshirani, Robert},
  title     = {Cross-Validation: What Does It Estimate and How Well Does It Do It?},
  journal   = {Journal of the American Statistical Association},
  year      = {2024},
  volume    = {119},
  number    = {546},
  pages     = {1434--1445},
  doi       = {10.1080/01621459.2023.2197686}
}

@article{beam2018,
  author    = {Beam, Andrew L. and Kohane, Isaac S.},
  title     = {Big Data and Machine Learning in Health Care},
  journal   = {JAMA},
  year      = {2018},
  volume    = {319},
  number    = {13},
  pages     = {1317--1318},
  doi       = {10.1001/jama.2017.18391}
}

@article{bengio2004,
  author    = {Bengio, Yoshua and Grandvalet, Yves},
  title     = {No Unbiased Estimator of the Variance of {K}-Fold Cross-Validation},
  journal   = {Journal of Machine Learning Research},
  year      = {2004},
  volume    = {5},
  pages     = {1089--1105}
}

@book{bischl2024,
  author    = {Bischl, Bernd and Sonabend, Raphael and Kotthoff, Lars and Lang, Michel},
  title     = {Applied Machine Learning Using mlr3 in {R}},
  publisher = {CRC Press},
  year      = {2024}
}

@article{cawley2010,
  author    = {Cawley, Gavin C. and Talbot, Nicola L. C.},
  title     = {On Over-Fitting in Model Selection and Subsequent Selection Bias in Performance Evaluation},
  journal   = {Journal of Machine Learning Research},
  year      = {2010},
  volume    = {11},
  pages     = {2079--2107}
}

@book{christen2012,
  author    = {Christen, Peter},
  title     = {Data Matching: Concepts and Techniques for Record Linkage, Entity Resolution, and Duplicate Detection},
  publisher = {Springer},
  year      = {2012},
  doi       = {10.1007/978-3-642-31164-2}
}

@article{gentleman2007,
  author    = {Gentleman, Robert and Lang, Duncan Temple},
  title     = {Statistical Analyses and Reproducible Research},
  journal   = {Journal of Computational and Graphical Statistics},
  year      = {2007},
  volume    = {16},
  number    = {1},
  pages     = {1--23},
  doi       = {10.1198/106186007X178663}
}

@article{golub1999,
  author    = {Golub, Todd R. and Slonim, Donna K. and Tamayo, Pablo and Huard, Christine and Gaasenbeek, Michelle and Mesirov, Jill P. and Coller, Hilary and Loh, Mignon L. and Downing, James R. and Caligiuri, Mark A. and Bloomfield, Clara D. and Lander, Eric S.},
  title     = {Molecular Classification of Cancer: Class Discovery and Class Prediction by Gene Expression Monitoring},
  journal   = {Science},
  year      = {1999},
  volume    = {286},
  number    = {5439},
  pages     = {531--537},
  doi       = {10.1126/science.286.5439.531}
}

@misc{hamdan2023,
  author    = {Hamdan, Sami and Love, Bradley C. and von dem Hagen, Elise A. H.},
  title     = {Confound-Leaking: Regression Leak in Machine Learning Applied to Neuroimaging},
  year      = {2023},
  eprint    = {2306.12952},
  archiveprefix = {arXiv},
  primaryclass  = {cs.LG},
  note      = {arXiv preprint}
}

@book{hastie2009,
  author    = {Hastie, Trevor and Tibshirani, Robert and Friedman, Jerome},
  title     = {The Elements of Statistical Learning: Data Mining, Inference, and Prediction},
  edition   = {2nd},
  publisher = {Springer},
  year      = {2009},
  doi       = {10.1007/978-0-387-84858-7}
}

@article{huber1964,
  author    = {Huber, Peter J.},
  title     = {Robust Estimation of a Location Parameter},
  journal   = {The Annals of Mathematical Statistics},
  year      = {1964},
  volume    = {35},
  number    = {1},
  pages     = {73--101},
  doi       = {10.1214/aoms/1177703732}
}

@book{huber2009,
  author    = {Huber, Peter J. and Ronchetti, Elvezio M.},
  title     = {Robust Statistics},
  edition   = {2nd},
  publisher = {Wiley},
  year      = {2009},
  doi       = {10.1002/9780470434697}
}

@article{kapoor2023,
  author    = {Kapoor, Sayash and Narayanan, Arvind},
  title     = {Leakage and the Reproducibility Crisis in Machine-Learning-Based Science},
  journal   = {Patterns},
  year      = {2023},
  volume    = {4},
  number    = {9},
  pages     = {100804},
  doi       = {10.1016/j.patter.2023.100804}
}

@article{kaufman2012,
  author    = {Kaufman, Shachar and Rosset, Saharon and Perlich, Claudia and Stitelman, Ori},
  title     = {Leakage in Data Mining: Formulation, Detection, and Avoidance},
  journal   = {ACM Transactions on Knowledge Discovery from Data},
  year      = {2012},
  volume    = {6},
  number    = {4},
  pages     = {1--21},
  doi       = {10.1145/2382577.2382579}
}

@inproceedings{kohavi1995,
  author    = {Kohavi, Ron},
  title     = {A Study of Cross-Validation and Bootstrap for Accuracy Estimation and Model Selection},
  booktitle = {Proceedings of the 14th International Joint Conference on Artificial Intelligence (IJCAI)},
  year      = {1995},
  volume    = {14},
  pages     = {1137--1145}
}

@misc{korkmaz2026,
  author        = {Korkmaz, Selcuk and G\"oks\"ul\"uk, Dincer and Karaismailo\u{g}lu, Eda},
  title         = {\textbf{fastml}: Guarded Resampling Workflows for Safer Automated Machine Learning in {R}},
  year          = {2026},
  eprint        = {2604.05225},
  archivePrefix = {arXiv},
  url           = {https://arxiv.org/abs/2604.05225}
}

@article{kuhn2008,
  author    = {Kuhn, Max},
  title     = {Building Predictive Models in {R} Using the \textbf{caret} Package},
  journal   = {Journal of Statistical Software},
  year      = {2008},
  volume    = {28},
  number    = {5},
  pages     = {1--26},
  doi       = {10.18637/jss.v028.i05}
}

@misc{kuhn2020,
  author    = {Kuhn, Max and Wickham, Hadley},
  title     = {Tidymodels: A Collection of Packages for Modeling and Machine Learning Using Tidyverse Principles},
  year      = {2020},
  url       = {https://www.tidymodels.org}
}

@article{lang2019,
  author    = {Lang, Michel and Binder, Martin and Richter, Jakob and Schratz, Patrick and Pfisterer, Florian and Coors, Stefan and Au, Quay and Casalicchio, Giuseppe and Kotthoff, Lars and Bischl, Bernd},
  title     = {\textbf{mlr3}: A Modern Object-Oriented Machine Learning Framework in {R}},
  journal   = {Journal of Open Source Software},
  year      = {2019},
  volume    = {4},
  number    = {44},
  pages     = {1903},
  doi       = {10.21105/joss.01903}
}

@book{leskovec2014,
  author    = {Leskovec, Jure and Rajaraman, Anand and Ullman, Jeffrey D.},
  title     = {Mining of Massive Datasets},
  edition   = {2nd},
  publisher = {Cambridge University Press},
  year      = {2014}
}

@article{moscovich2022,
  author    = {Moscovich, Amit and Rosset, Saharon},
  title     = {On the Cross-Validation Bias Due to Unsupervised Pre-Processing},
  journal   = {Journal of the Royal Statistical Society: Series B (Statistical Methodology)},
  year      = {2022},
  volume    = {84},
  number    = {4},
  pages     = {1474--1502},
  doi       = {10.1111/rssb.12537}
}

@article{phipson2010,
  author    = {Phipson, Belinda and Smyth, Gordon K.},
  title     = {Permutation P-values Should Never Be Zero: Calculating Exact P-values When Permutations Are Randomly Drawn},
  journal   = {Statistical Applications in Genetics and Molecular Biology},
  year      = {2010},
  volume    = {9},
  number    = {1},
  pages     = {Article 39},
  doi       = {10.2202/1544-6115.1585}
}

@article{peng2011,
  author    = {Peng, Roger D.},
  title     = {Reproducible Research in Computational Science},
  journal   = {Science},
  year      = {2011},
  volume    = {334},
  number    = {6060},
  pages     = {1226--1227},
  doi       = {10.1126/science.1213847}
}

@Manual{R,
  title     = {{R}: A Language and Environment for Statistical Computing},
  author    = {{R Core Team}},
  organization = {R Foundation for Statistical Computing},
  address   = {Vienna, Austria},
  year      = {2025},
  url       = {https://www.R-project.org/}
}

@article{roberts2017,
  author    = {Roberts, David R. and Bahn, Volker and Ciuti, Simone and Boyce, Mark S. and Elith, Jane and Guillera-Arroita, Gurutzeta and Hauenstein, Severin and Lahoz-Monfort, Jos{\'e} J. and Schr{\"o}der, Boris and Thuiller, Wilfried and Warton, David I. and Wintle, Brendan A. and Hartig, Florian and Dormann, Carsten F.},
  title     = {Cross-Validation Strategies for Data with Temporal, Spatial, Hierarchical, or Phylogenetic Structure},
  journal   = {Ecography},
  year      = {2017},
  volume    = {40},
  number    = {8},
  pages     = {913--929},
  doi       = {10.1111/ecog.02881}
}

@article{rosenblatt2024,
  author    = {Rosenblatt, Jonathan D. and Ben Simhon, Yaron and Gal, Sagiv and Gilron, Ro{'}ee},
  title     = {Practicalities of Resampling Methods for Brain Imaging: What Every Researcher Should Know},
  journal   = {NeuroImage},
  year      = {2024},
  volume    = {287},
  pages     = {120520},
  doi       = {10.1016/j.neuroimage.2024.120520}
}

@article{sandve2013,
  author    = {Sandve, Geir Kjetil and Nekrutenko, Anton and Taylor, James and Hovig, Eivind},
  title     = {Ten Simple Rules for Reproducible Computational Research},
  journal   = {PLOS Computational Biology},
  year      = {2013},
  volume    = {9},
  number    = {10},
  pages     = {e1003285},
  doi       = {10.1371/journal.pcbi.1003285}
}

@inproceedings{sendak2020,
  author    = {Sendak, Mark and Elish, Madeleine Clare and Gao, Michael and Futoma, Joseph and Ratliff, William and Nichols, Marshall and Bedoya, Armando and Balu, Suresh and O'Brien, Cara},
  title     = {The Human Body Is a Black Box: Supporting Clinical Decision-Making with Deep Learning},
  booktitle = {Proceedings of the Conference on Fairness, Accountability, and Transparency (FAT* '20)},
  year      = {2020},
  pages     = {99--109},
  doi       = {10.1145/3351095.3372827}
}

@article{simon2003,
  author    = {Simon, Richard and Radmacher, Michael D. and Dobbin, Kevin and McShane, Lisa M.},
  title     = {Pitfalls in the Use of {DNA} Microarray Data for Diagnostic and Prognostic Classification},
  journal   = {Journal of the National Cancer Institute},
  year      = {2003},
  volume    = {95},
  number    = {1},
  pages     = {14--18},
  doi       = {10.1093/jnci/95.1.14}
}

@article{soneson2014,
  author    = {Soneson, Charlotte and Delorenzi, Mauro},
  title     = {Batch Effect Confounding Leads to Strong Bias in Performance Estimates Obtained by Cross-Validation},
  journal   = {PLOS ONE},
  year      = {2014},
  volume    = {9},
  number    = {6},
  pages     = {e100335},
  doi       = {10.1371/journal.pone.0100335}
}

@article{stone1974,
  author    = {Stone, Mervyn},
  title     = {Cross-Validatory Choice and Assessment of Statistical Predictions},
  journal   = {Journal of the Royal Statistical Society: Series B (Methodological)},
  year      = {1974},
  volume    = {36},
  number    = {2},
  pages     = {111--147},
  doi       = {10.1111/j.2517-6161.1974.tb00994.x}
}

@article{subramanian2013,
  author    = {Subramanian, Janani and Simon, Richard},
  title     = {Overfitting in Prediction Models -- Is It a Problem Only in High Dimensions?},
  journal   = {Contemporary Clinical Trials},
  year      = {2013},
  volume    = {36},
  number    = {2},
  pages     = {636--641},
  doi       = {10.1016/j.cct.2013.06.011}
}

@article{tarca2007,
  author    = {Tarca, Adi L. and Carey, Vincent J. and Chen, Xue-wen and Romero, Roberto and Drăghici, Sorin},
  title     = {Machine Learning and Its Applications to Biology},
  journal   = {PLOS Computational Biology},
  year      = {2007},
  volume    = {3},
  number    = {6},
  pages     = {e116},
  doi       = {10.1371/journal.pcbi.0030116}
}

@article{valavi2019,
  author    = {Valavi, Roozbeh and Elith, Jane and Lahoz-Monfort, Jos{\'e} J. and Guillera-Arroita, Gurutzeta},
  title     = {\textbf{blockCV}: An {R} Package for Generating Spatially or Environmentally Separated Folds for \textit{k}-Fold Cross-Validation of Species Distribution Models},
  journal   = {Methods in Ecology and Evolution},
  year      = {2019},
  volume    = {10},
  number    = {2},
  pages     = {225--232},
  doi       = {10.1111/2041-210X.13107}
}

@article{vandemortel2025,
  author    = {van de Mortel, Laurens A. and van Wingen, Guido A.},
  title     = {Data Leakage Inflates Prediction Performance in Connectome-Based Machine Learning Models},
  journal   = {Nature Communications},
  year      = {2025},
  volume    = {16},
  pages     = {1106},
  doi       = {10.1038/s41467-025-56082-4}
}

@article{varma2006,
  author    = {Varma, Sudhir and Simon, Richard},
  title     = {Bias in Error Estimation When Using Cross-Validation for Model Selection},
  journal   = {BMC Bioinformatics},
  year      = {2006},
  volume    = {7},
  pages     = {91},
  doi       = {10.1186/1471-2105-7-91}
}

@article{wenger2012,
  author    = {Wenger, Seth J. and Olden, Julian D.},
  title     = {Assessing Transferability of Ecological Models: An Underappreciated Aspect of Statistical Validation},
  journal   = {Methods in Ecology and Evolution},
  year      = {2012},
  volume    = {3},
  number    = {2},
  pages     = {260--267},
  doi       = {10.1111/j.2041-210X.2011.00170.x}
}

@techreport{winkler1999,
  author    = {Winkler, William E.},
  title     = {The State of Record Linkage and Current Research Problems},
  institution = {U.S. Bureau of the Census},
  year      = {1999},
  type      = {Statistical Research Report Series},
  number    = {RR99/04}
}

\end{document}